\documentclass[a4paper,11pt]{article}
\pdfoutput=1
\usepackage{jcappub}
\usepackage{graphicx}
\usepackage{bm}
\usepackage{amssymb,amsmath,bm,dsfont}
\usepackage{color}
\usepackage[dvipsnames]{xcolor}
\usepackage[utf8]{inputenc}
\usepackage{balance}
\usepackage{enumitem}
\usepackage{lipsum}
\usepackage[caption=false]{subfig}

\newcommand{\nv}{\hat{\boldsymbol{\theta}}}
\newcommand{\summ}[1]{\sum_{\bf #1}\Delta^2 #1\,}
\newcommand{\Ylm}[3]{{\sf E}^{#1}_{\bf #2}({\bf #3})}
\newcommand{\wtj}[6]{\left(\begin{array}{ccc} #1 & #2 & #3\\#4 & #5 & #6\end{array} \right)}

\title{\boldmath Disconnected pseudo-$C_\ell$ covariances for projected large-scale structure data}

\author[a,b]{Carlos Garc\'{i}a-Garc\'{i}a}
\author[c]{David Alonso}
\author[c]{Emilio Bellini}

\affiliation[a]{Instituto de Física Fundamental, Consejo Superior de Investigaciones Científicas, c/. Serrano 123, E–28006, Madrid, Spain}
\affiliation[b]{Institut de Ci\`{e}ncies del Cosmos (UB–IEEC), c/. Martí i Franqués 1, E–08028, Barcelona, Spain}
\affiliation[c]{Oxford Astrophysics, Department of Physics, Keble Road, Oxford, OX1 3RH, UK}

\emailAdd{carlosgarcia@iff.csic.es}
\emailAdd{david.alonso@physics.ox.ac.uk}
\emailAdd{emilio.bellini@physics.ox.ac.uk}

\abstract{The disconnected part of the power spectrum covariance matrix (also
  known as the ``Gaussian'' covariance) is the dominant contribution on large
  scales for galaxy clustering and weak lensing datasets. The presence of a
  complicated sky mask causes non-trivial correlations between different
  Fourier/harmonic modes, which must be accurately characterized in order to
  obtain reliable cosmological constraints. This is particularly relevant for
  galaxy survey data. Unfortunately, an exact calculation of these
  correlations involves $O(\ell_{\rm max}^6)$ operations that become
  computationally impractical very quickly. We present an implementation of
  approximate methods to estimate the Gaussian covariance matrix of power
  spectra involving spin-0 and spin-2 flat- and curved-sky fields, expanding
  on existing algorithms developed in the context of CMB analyses.
  These methods achieve an $O(\ell_{\rm max}^3)$
  scaling, which makes the computation of the covariance matrix as fast as the
  computation of the power spectrum itself. We quantify the accuracy of these
  methods on large-scale structure and weak lensing data, making use of a
  large number of Gaussian but otherwise realistic simulations. We show that,
  using the approximate covariance matrix, we are able to recover the true
  posterior distribution of cosmological parameters to high accuracy. We also
  quantify the shortcomings of these methods, which become unreliable on the
  very largest scales, as well as for covariance matrix elements involving
  cosmic shear $B$ modes. The algorithms presented here are implemented in the
  public code {\tt NaMaster} \url{https://github.com/LSSTDESC/NaMaster}.}

\begin{document}
  \maketitle
  \flushbottom

  \section{Introduction}\label{sec:intro}
    The two-point correlation of different fields projected on the celestial
    sphere is one of the most common observables used in the analysis of large
    datasets in astrophysics, from studies of the Cosmic Microwave Background
    \cite{Gorski:1994ye,Gorski:1994uu,1995PhRvL..74.4369B,Gorski:1996cf,Gorski:1996ti,1997PhRvD..55.5895T,1998PhRvD..57.2117B,Wandelt:2000av,2001PhRvD..64f3001T,2002ApJ...567....2H,2003PhRvD..67b3001W}
    to large-scale structure and weak lensing surveys
    \cite{2000MNRAS.317L..23H,2001ApJ...555..547H,2001MNRAS.325.1603E,2011arXiv1112.5723H,2016MNRAS.456.1508K,2018MNRAS.476.1050B,2018arXiv181208182X,2019MNRAS.tmp.1446C,2019PASJ...71...43H}.
    Using these two-point functions, one achieves a high level of data
    compression (with respect to the size of the raw datasets -- time-ordered
    data, images or catalogs). They can also be directly used to constrain
    cosmological and astrophysical parameters assuming that one can model
    their likelihood. This is usually done by assuming that the two-point
    functions are Gaussianly distributed, which is often a good approximation
    due to the central limit theorem
    \cite{2008PhRvD..77j3013H,2018MNRAS.477.4879S}. In this case, the only
    obstacle that remains is being able to estimate the covariance matrix of a
    set of two-point correlators. Since the form of this covariance directly
    affects the posterior parameter uncertainties, a precise determination of
    it is of paramount importance. In large-scale structure experiments, this
    has often been resolved by making use of one's own data through resampling
    techniques
    \cite{tukey1958,1984MNRAS.210P..19B,1993ApJ...406L..47H,2002ApJ...571..172Z,2016arXiv160600233E},
    or by generating a large number of mock realizations
    \cite{2013MNRAS.428.1036M,2018MNRAS.479...94A,2019MNRAS.485.2806B}. With
    the advent of the larger current and future surveys, the increasing size
    of the data vector and of the volume to be simulated has made this
    solution impractical, and fully analytical and hybrid estimators are now
    being used.
  
    The problem of producing accurate analytical estimates of the covariance matrix for large-scale structure data has seen significant progress in the last few years \cite{2001ApJ...554...56C,2009ApJ...701..945S,2009MNRAS.395.2065T,2013PhRvD..87l3504T,2014MNRAS.441.2456T,2016PhRvD..94f3533P}. As described in \cite{2018JCAP...10..053B}, the covariance matrix recieves three main contributions:
    \begin{itemize}
      \item {\bf Gaussian covariance:} this is the contribution to the covariance from the disconnected part of the trispectrum of the different fields involved (also called the ``disconnected'' covariance). In simpler terms, this is the covariance matrix one would obtain if all fields involved were Gaussianly distributed.
      \item {\bf Connected non-Gaussian covariance:} this is the contribution from the connected trispectrum (which would vanish if all fields were Gaussianly distributed).
      \item {\bf Super-sample covariance:} this is the additional coupling between different scales induced by density fluctuations on scales larger than the volume mapped. This term also vanishes for Gaussian fields.
    \end{itemize}
  
    On most scales relevant for cosmological studies, the Gaussian contribution dominates the error budget, although the connected and super-sample terms cannot be neglected \cite{2018JCAP...10..053B}. The Gaussian contribution is trivial to compute for fields observed over the full sky:
    \begin{equation}\label{eq:cov_naive}
      {\rm Cov}\left(C^{ab}_\ell,C^{cd}_\ell\right)=\delta^K_{\ell\ell'}\frac{C^{ac}_\ell C^{bd}_\ell+C^{ad}_\ell C^{bc}_\ell}{2\ell+1},
    \end{equation}
    where $C^{xy}_\ell$ is the angular power spectrum between two maps $x$ and $y$ on multipole $\ell$. Unfortunately, the presence of a sky mask in general induces non-trivial couplings between different $\ell$s, which must be accurately estimated in order to produce unbiased evaluations of the parameter likelihood \cite{2018MNRAS.479.4998T}.
  
    In this paper, we will present and generalize methods developed in the context of CMB experiments to account for the impact of survey geometry on the Gaussian part of the power spectrum covariance matrix \cite{2004MNRAS.349..603E,2005MNRAS.360.1262B,2005MNRAS.360..509C,Efstathiou:2006eb,2014A&A...571A..15P,2016A&A...594A..11P,2017A&A...602A..41C,2019arXiv190712875P}\footnote{See also \cite{2019JCAP...01..016L} for a similar application to the problem of 3D power spectrum covariances.}, and will study in detail the performance of these methods for large-scale structure and weak lensing datasets. We have also implemented these approximations in the public code {\tt NaMaster}\footnote{\url{https://github.com/LSSTDESC/NaMaster}} \cite{2019MNRAS.484.4127A}, making the computation of accurate Gaussian covariance matrices significantly simpler for the community.

    The paper is structured as follows: Section \ref{sec:theory} presents the methods and approximations used to calculate accurate covariances. In Section \ref{sec:results} we test the methods against Gaussian simulations and study their performance as well as their impact on the final cosmological parameter estimation. We then summarize our results and conclude in Section \ref{sec:discussion}. Appendix \ref{app:flat} presents the performance of these methods in the flat-sky approximation, and we provide technical details of the software implementation in Appendix \ref{app:namaster}.

  \section{Analytical Gaussian covariances}\label{sec:theory}
    \subsection{Preliminaries}\label{ssec:theory.prelim}
      We will deal with spin-0 and spin-2 fields defined on a 2-dimensional space. In two dimensions, spin-$s$ fields in general have two components ${\bf a}({\bf x})=(a_1({\bf x}),a_2({\bf x}))$\footnote{E.g. for CMB polarization, a spin-2 field, these components are the Stokes parameters $(Q,U)$, while for cosmic shear the two components are usually labeled $(\gamma_1,\gamma_2)$.}. Forming a complex number from these components, $a_1+ia_2$, spin-$s$ fields transform, under a coordinate rotation with angle $\psi$, as $a_1+ia_2\rightarrow (a_1+ia_2)\exp(i\,s\psi)$. Thus, spin-0 fields are invariant under rotations, and are usually expressed as real-valued fields with a single component.
      
      Given a field ${\bf a}({\bf x})$, with 1 (spin-0) or 2 components (spin-2), defined on the coordinates ${\bf x}$, we define its generalized Fourier coefficients as
      \begin{equation}
        {\bf a}_{\bf k}=\summ{x}\,\Ylm{\dag}{k}{x}\,{\bf a}({\bf x}),
      \end{equation}
      where the operator $\summ{x}$ denotes an integral or sum over all values of the coordinates ${\bf x}$, and $\Ylm{\dag}{k}{x}$ are a set of orthogonal functions. We will also assume that the $\Ylm{\dag}{k}{x}$ are a \emph{complete} set of basis functions, in which case:
      \begin{align}
        &\summ{x}\,\Ylm{\dag}{k}{x}\Ylm{}{l}{x}=\mathds{1}\Delta^x({\bf k},{\bf l}),\\
        &\summ{k}\,\Ylm{}{k}{x}\Ylm{\dag}{k}{y}=\mathds{1}\Delta^k({\bf x},{\bf y}),
      \end{align}
      where $\summ{k}$ denotes an integral over all possible generalized Fourier coefficients ${\bf k}$, and $\Delta^x$ and $\Delta^k$ are generalized delta functions, defined through their action on functions of ${\bf x}$ or ${\bf k}$:
      \begin{align}
        &\summ{k}\,f({\bf k})\Delta^x({\bf k},{\bf l})\equiv f({\bf l}),\\
        &\summ{x}\,f({\bf x})\Delta^k({\bf x},{\bf y})\equiv f({\bf y}).
      \end{align}
      For a spin-$s$ quantity, $\Ylm{}{l}{x}$ can be written in terms of two spin-raising and spin-lowering operators, $\eth$ and $\bar{\eth}$, and a set of scalar orthogonal functions $q({\bf l},{\bf x})$ as:
      \begin{equation}
        \Ylm{}{l}{x}=-\frac{\beta_{\ell,s}}{2}\left(
        \begin{array}{cc}
          \eth^s+\bar{\eth}^s & i(\eth^s-\bar{\eth}^s)\\
          -i(\eth^s-\bar{\eth}^s) & \eth^s+\bar{\eth}^s
        \end{array}\right) q({\bf l},{\bf x}),
      \end{equation}
      where $\beta_{\ell,s}$ is a normalization factor defined in Table \ref{tab:notation}.
    
      Finally, we will assume that all fields are Gaussian stochastic fields that are additionally statistically isotropic. As a consequence of the latter, different generalized Fourier modes are uncorrelated:
      \begin{equation}\label{eq:iso}
        \langle {\bf a}_{\bf k}{\bf b}^\dag_{\bf l}\rangle\equiv {\sf C}^{ab}_\ell\,K\,\Delta^x({\bf k},{\bf l}),
      \end{equation}
      where $K$ is a volume factor (see below) and ${\sf C}^{ab}_k$ is the power spectrum. Defined this way, the power spectrum is a matrix, with elements
      \begin{equation}
        \left\langle a^\alpha_{\bf k}\left(b^\beta_{\bf l}\right)^*\right\rangle\equiv \left({\sf C}^{ab}_\ell\right)_{\alpha\beta}\,K\,\Delta^x({\bf k},{\bf l}),
      \end{equation}
      where $a^\alpha$ is the $\alpha$-th element of field ${\bf a}$. It will often be useful in what follows to think of ${\sf C}^{ab}_\ell$ as a 1-dimensional vector that we will denote by ${\rm vec}({\sf C}^{ab}_\ell)$. To do so, we simply map the two indices $(\alpha,\beta)$ into a single number $A$, such that ${\rm vec}({\sf C}^{ab}_\ell)_A = ({\sf C}^{ab}_\ell)_{\alpha\beta}$.
    
      \begin{table*}
        \centering
        \begin{tabular}{|c|c|c|}
          \hline
          Symbol & Curved sky & Flat sky (continuum $\rightarrow$ discretized) \\
          \hline
          \(\displaystyle {\bf l} \) & \(\displaystyle (\ell,m) \) & \(\displaystyle (l_x,l_y) \) \\
          \(\displaystyle \summ{l} \) & \(\displaystyle \sum_{\ell=0}^{\infty} \sum_{m=-\ell}^{\ell} \) & \(\displaystyle \int \frac{dl^2}{2\pi} \rightarrow \sum_{\bf l}\frac{2\pi}{L_xL_y}\) \\
          \(\displaystyle \Delta^x({\bf l},{\bf l}')\) & \(\displaystyle \delta^K_{\ell\ell'}\delta^K_{mm'}\) & \(\displaystyle 2\pi\,\delta^D({\bf l}-{\bf l}')\rightarrow \delta^K_{l_xl'_x}\delta^K_{l_yl'_y}\frac{L_xL_y}{(2\pi)^2}2\pi\) \\
          \(\displaystyle {\bf x}\) & \(\displaystyle \nv\equiv(\theta,\varphi) \) & \(\displaystyle (x,y)\) \\
          \(\displaystyle \summ{x} \) & \(\displaystyle \int_0^{\phi} d\phi \int_{-1}^{1} d(\cos\theta)\) & \(\displaystyle \int \frac{dx^2}{2\pi}\rightarrow\sum_{\bf x}\frac{L_xL_y}{2\pi N_xN_y}\) \\
          \(\displaystyle \Delta^k({\bf x},{\bf y})\) & \(\displaystyle \delta^D(\cos\theta-\cos\theta')\delta^D(\varphi-\varphi')\) & \(\displaystyle 2\pi\,\delta^D({\bf x}-{\bf y})\rightarrow\delta^K_{xx'}\delta^K_{yy'}\frac{N_xN_y}{L_xL_y}2\pi\) \\
          \(\displaystyle q({\bf l},{\bf x})\) & \(\displaystyle Y_{\ell m}(\nv)\) & \(\displaystyle e^{i{\bf l}\cdot{\bf x}}\) \\
          \(\displaystyle \eth\,_sf\) & \(\displaystyle -(\sin\theta)^s\left(\partial_\theta+i\frac{\partial_\varphi}{\sin\theta}\right)(\sin\theta)^{-s}\,_sf(\nv)\) & \(\displaystyle (\partial_x-i\partial_y)\,_sf\) \\
          \(\displaystyle \bar{\eth}\,_sf\) & \(\displaystyle -(\sin\theta)^{-s}\left(\partial_\theta-i\frac{\partial_\varphi}{\sin\theta}\right)(\sin\theta)^s\,_sf(\nv)\) & \(\displaystyle (\partial_x+i\partial_y)\,_sf\) \\
          \(\displaystyle \beta_{\ell,s}\) & \(\displaystyle \sqrt{\frac{(\ell-s)!}{(\ell+s)!}}\) & \(\displaystyle \ell^{-s}\) \\
          \(\displaystyle K \) & \(\displaystyle 1 \) & \(\displaystyle (2\pi)^{-1}\) \\
          \hline
        \end{tabular}
        \caption{Lookup table describing the generalized notation introduced in Section \ref{ssec:theory.prelim} for quantities defined on the sphere (second column) and on the flat 2D plane (third column). For the flat-sky case, we also provide expressions for a discretized, finite 2D plane with periodic boundary conditions. In this case, the map has dimensions $(L_x,L_y)$ subdivided into $(N_x,N_y)$ equi-spaced pixels in $(x,y)$. $\delta^D$ and $\delta^K$ are the Dirac and Kronecker delta functions respectively.}\label{tab:notation}
      \end{table*}

      All the functions and operators above can be specialized to fields defined on the sphere or the 2D plane (flat sky approximation) as described in Table \ref{tab:notation}.

    \subsection{The pseudo-$C_\ell$ method}\label{ssec:theory.pcl}
      This section provides a very brief introduction to the pseudo-$C_\ell$ power spectrum estimator. Further details can be found in e.g. \cite{2002ApJ...567....2H,2005MNRAS.360.1262B,2019MNRAS.484.4127A}. 
      In any practical situation we do not have access to maps of a given field ${\bf a}$ over the full sky, but rather to a weighted or masked version of them
      \begin{equation}
        \tilde{\bf a}({\bf x})\equiv w_a({\bf x}){\bf a}({\bf x}),
        \label{eq:mask}
      \end{equation}
      where $w_a$ is commonly called the ``mask''. Due to the convolution theorem, the generalized Fourier coefficients of the masked field will be a convolution of the mask and true field coefficients:
      \begin{align}\nonumber
        \tilde{\bf a}_{\bf l}=&\summ{k}\left[\summ{x}\,w^a({\bf x})\,\Ylm{\dag}{l}{x}\,\Ylm{}{k}{x}\right]{\bf a}_{\bf k}\\
                        \equiv&\summ{k}\,^a{\sf M}_{{\bf l}{\bf k}}\,{\bf a}_{\bf k},
      \end{align}
      where we have defined the mode-coupling coefficients $\,^a{\sf M}_{{\bf l}{\bf k}}$ in the second line.
    
      Correlating the generalized Fourier coefficients of two masked fields therefore yields a mode-coupled version of their true underlying power spectrum:
      \begin{align}\nonumber
        \left\langle\tilde{\bf a}_{\bf l}\tilde{\bf b}^\dag_{\bf l}\right\rangle=&\summ{k}\summ{q}\,^a{\sf M}_{{\bf l}{\bf k}}\left\langle{\bf a}_{\bf k}{\bf b}^\dag_{\bf q}\right\rangle\,^b{\sf M}^\dag_{{\bf l}{\bf q}}\\
                                                                              =&K\,\summ{k}\,^a{\sf M}_{{\bf l}{\bf k}}{\sf C}^{ab}_k\,^b{\sf M}^\dag_{{\bf l}{\bf k}} \label{eq:M}                                                                             
      \end{align}
    
      The pseudo-$C_\ell$ estimator then proceeds in two steps:
      \begin{enumerate}
        \item We first bin different ${\bf l}$ modes into sets of them called bandpowers (typically bands of similar $\ell$ or annuli of flat-sky Fourier modes spanning a range of radii). Let us denote a given bandpower by its index $q$. We must note that it is more appropriate to use bandpower-averaged spectra when the underlying spectrum does not vary much within each $\ell$ bin. When this is not the case, it is often useful to apply $\ell$-weights (such as $D_\ell\equiv\ell(\ell+1)C_\ell/(2\pi)$). The large-scale structure spectra discussed here are sufficiently flat that the binning used is appropriate. The binned pseudo-power spectrum is:
        \begin{equation}\label{eq:bandpowers}
          \tilde{\sf C}^{ab}_q=\sum_{{\bf l}\in q} B_q^{\bf l}\,\tilde{\bf a}_{\bf l}\tilde{\bf b}^\dag_{\bf l},
        \end{equation}
        where the bandpower weights are normalized such that $\sum_{{\bf l}\in
          q}B_q^{\bf l}=(K\Delta^x({\bf 0}))^{-1}$.
        \item Then, the correlation between bandpowers induced by the mode-coupling coefficients is partially reversed by multiplying $\tilde{\sf C}^{ab}_q$ by the so-called binned ``mode-coupling matrix'' $\mathcal{M}$, giving the final estimator
        \begin{equation}
          {\rm vec}\left(\hat{\sf C}^{ab}_q\right) = \sum_{q'} \left(\mathcal{M}^{-1}\right)_{qq'}\,{\rm vec}\left(\tilde{\sf C}^{ab}_{q'}\right).
        \end{equation}

        The main advantage of the pseudo-$C_\ell$ estimator is that the mode-coupling matrix $\mathcal{M}$ is directly related to the coupling coefficients $\,^a{\sf M}_{{\bf l}{\bf l}'}$, and can be computed analytically making use of methods that scale like $\ell_{\rm max}^3$ (see e.g. \cite{2002ApJ...567....2H}).
      \end{enumerate}
    
      For completeness, the mode-coupling matrices for flat-sky and curved-sky fields are given by \cite{2019MNRAS.484.4127A}:
      \begin{itemize}
        \item {\bf Curved sky.} After averaging over the harmonic number $m$, the mode-coupling matrices are:
        \begin{equation}\label{eq:mcms}
          \left\langle\frac{1}{2\ell+1}\sum_{m=-\ell}^\ell {\rm vec}\left[\tilde{\bf a}_{\bf l}\tilde{\bf b}^\dag_{\bf l}\right]\right\rangle=\sum_{\ell'}{\sf M}^{s_as_b}_{\ell\ell'} {\rm vec}\left[{\sf C}^{ab}_{\ell'}\right],
        \end{equation}
        with
        \begin{align}
          &{\sf M}^{00}_{\ell\ell'}=(2\ell'+1)\,\Xi^{00}_{\ell\ell'}(w_a,w_b),\hspace{12pt}
          {\sf M}^{02}_{\ell\ell'}=(2\ell'+1)\,\Xi^{0+}_{\ell\ell'}(w_a,w_b)\,\mathds{1},\\
          &{\sf M}^{22}_{\ell\ell'}=(2\ell'+1)
          \left(
          \begin{array}{cccc}
            \Xi^{++}_{\ell\ell'} & 0 & 0 & \Xi^{--}_{\ell\ell'} \\
            0 & \Xi^{++}_{\ell\ell'} & -\Xi^{--}_{\ell\ell'} & 0 \\
            0 & -\Xi^{--}_{\ell\ell'} & \Xi^{++}_{\ell\ell'} & 0 \\
            \Xi^{--}_{\ell\ell'} & 0 & 0 & \Xi^{++}_{\ell\ell'}
          \end{array}\right),
        \end{align}
        where
        \begin{align}\label{eq:coeff_mcm}
          &\Xi^{00}_{\ell\ell'}(w,v)  \equiv\sum_{\ell''}\frac{P^{wv}_{\ell''}}{4\pi}\wtj{\ell}{\ell'}{\ell''}{0}{0}{0}^2\\
          &\Xi^{0+}_{\ell\ell'}(w,v)  \equiv\sum_{\ell''}\frac{P^{wv}_{\ell''}}{4\pi}\wtj{\ell}{\ell'}{\ell''}{0}{0}{0}\wtj{\ell}{\ell'}{\ell''}{2}{-2}{0}\\
          &\Xi^{\pm\pm}_{\ell\ell'}(w,v) \equiv\sum_{\ell''}\frac{P^{wv}_{\ell''}}{4\pi}\wtj{\ell}{\ell'}{\ell''}{2}{-2}{0}^2\frac{1\pm(-1)^{\ell+\ell'+\ell''}}{2}.
        \end{align}
        Here the 2-by-3 matrix-like quantities are the Wigner 3-$j$ symbols, and
        \begin{equation}
          P^{vw}_\ell\equiv\sum_{m=-\ell}^\ell v_{\ell m}w^*_{\ell m}.
        \end{equation}
        \item {\bf Flat sky.} In this case the averaging over the Fourier-space azimuth happens while binning into bandpowers, and therefore the unbinned mode-coupling matrix is defined before binning. Assuming flat bandpowers, such that $B_q^{\bf l}=(2\pi)^2/(L_xL_yN_q)$, where $N_q$ is the number of Fourier-space modes in the $q$-th bandpower:
        \begin{equation}
          \left\langle{\rm vec}\left[\tilde{\sf C}^{ab}_q\right]\right\rangle=\sum_{{\bf l}\in q}\frac{1}{N_q} \sum_{\bf k}{\sf M}^{s_as_b}_{{\bf l}{\bf k}}\,{\rm vec}\left[{\sf C}^{ab}_k\right]
        \end{equation}
        with
        \begin{align}
          {\sf M}^{00}_{{\bf l}\,{\bf k}}&\equiv \bar{\Xi}^{00}_{{\bf l}{\bf k}},\\
          {\sf M}^{02}_{{\bf l}\,{\bf k}}&\equiv 
          \left(
          \begin{array}{cc}
            \bar{\Xi}^{0+}_{{\bf l}{\bf k}} & -\bar{\Xi}^{0-}_{{\bf l}{\bf k}} \\
            \bar{\Xi}^{0-}_{{\bf l}{\bf k}} &  \bar{\Xi}^{0+}_{{\bf l}{\bf k}}
          \end{array}\right)\\\label{eq:m22_flat}
          {\sf M}^{22}_{{\bf l}\,{\bf k}}&\equiv 
          \left(
          \begin{array}{cccc}
            \bar{\Xi}^{++}_{{\bf l}{\bf k}} & -\bar{\Xi}^{+-}_{{\bf l}{\bf k}} & -\bar{\Xi}^{+-}_{{\bf l}{\bf k}} &  \bar{\Xi}^{--}_{{\bf l}{\bf k}} \\
            \bar{\Xi}^{+-}_{{\bf l}{\bf k}} &  \bar{\Xi}^{++}_{{\bf l}{\bf k}} & -\bar{\Xi}^{--}_{{\bf l}{\bf k}} & -\bar{\Xi}^{+-}_{{\bf l}{\bf k}} \\
            \bar{\Xi}^{+-}_{{\bf l}{\bf k}} & -\bar{\Xi}^{--}_{{\bf l}{\bf k}} &  \bar{\Xi}^{++}_{{\bf l}{\bf k}} & -\bar{\Xi}^{+-}_{{\bf l}{\bf k}} \\
            \bar{\Xi}^{--}_{{\bf l}{\bf k}} &  \bar{\Xi}^{+-}_{{\bf l}{\bf k}} &  \bar{\Xi}^{+-}_{{\bf l}{\bf k}} & \bar{\Xi}^{++}_{{\bf l}{\bf k}}
          \end{array}\right),
        \end{align}
        where
        \begin{align}\label{eq:coeff_mcm_flat}
          &\bar{\Xi}^{00}_{{\bf l}{\bf k}}=\left(\frac{2\pi}{L_xL_y}\right)^2 (w_a)_{{\bf l}-{\bf k}}(w_b)^*_{{\bf l}-{\bf k}},\\
          &\bar{\Xi}^{0+}_{{\bf l}{\bf k}}=\left(\frac{2\pi}{L_xL_y}\right)^2 (w_a)_{{\bf l}-{\bf k}}(w_b)^*_{{\bf l}-{\bf k}}\cos2\Delta\varphi,\\
          &\bar{\Xi}^{0-}_{{\bf l}{\bf k}}=\left(\frac{2\pi}{L_xL_y}\right)^2 (w_a)_{{\bf l}-{\bf k}}(w_b)^*_{{\bf l}-{\bf k}}\sin2\Delta\varphi,\\
          &\bar{\Xi}^{++}_{{\bf l}{\bf k}}=\left(\frac{2\pi}{L_xL_y}\right)^2 (w_a)_{{\bf l}-{\bf k}}(w_b)^*_{{\bf l}-{\bf k}}\cos^22\Delta\varphi,\\
          &\bar{\Xi}^{+-}_{{\bf l}{\bf k}}=\left(\frac{2\pi}{L_xL_y}\right)^2 (w_a)_{{\bf l}-{\bf k}}(w_b)^*_{{\bf l}-{\bf k}}\cos2\Delta\varphi\,\sin2\Delta\varphi,\\
          &\bar{\Xi}^{--}_{{\bf l}{\bf k}}=\left(\frac{2\pi}{L_xL_y}\right)^2 (w_a)_{{\bf l}-{\bf k}}(w_b)^*_{{\bf l}-{\bf k}}\sin^22\Delta\varphi,
        \end{align}
        and $\Delta\varphi$ is the relative angle between ${\bf l}$ and ${\bf k}$.
      \end{itemize}

      Before we move on to covariances, it is worth considering the case of
      unmasked field (i.e $w^a({\bf x})=1$ everywhere). In this case $\,^a{\sf
        M}_{{\bf l}{\bf l}'}=\mathds{1}\Delta^x({\bf l},{\bf l}')$, and therefore different modes are uncorrelated (as should have been obvious). In a non-ideal case where the mask is still sufficiently well behaved (i.e. masks without too much small-scale structure), we can still expect the coupling coefficients $\,^a{\sf M}_{{\bf l}{\bf l}'}$ to be sharply peaked around ${\bf l}={\bf l}'$.
  
    \subsection{Covariance matrices}\label{ssec:theory.pclcov}
      So far we have not assumed anything about the statistics of the fields, other than the fact that they are isotropic (Eq. \ref{eq:iso}). This section presents a method to estimate the disconnected part of the power spectrum covariance for the pseudo-$C_\ell$ estimator.
    
      Let $A$ and $F$ be the vector indices corresponding to the pairs of
      field indices $(\alpha,\beta)$ and $(\phi,\gamma)$, respectively, and let us start by considering the covariance
      \begin{equation}
        \Sigma^{AF}_{{\bf l}{\bf l}'}\equiv\left\langle\tilde{a}^\alpha_{\bf l}\tilde{b}^{\beta *}_{\bf l}\tilde{f}^\phi_{{\bf l}'}\tilde{g}^{\gamma *}_{{\bf l}'}\right\rangle-\left\langle\tilde{a}^\alpha_{\bf l}\tilde{b}^{\beta *}_{\bf l}\right\rangle\left\langle\tilde{f}^\phi_{{\bf l}'}\tilde{g}^{\gamma *}_{{\bf l}'}\right\rangle
      \end{equation}
      The covariance of the binned bandpowers $\tilde{\sf C}$ can then be computed as
      \begin{equation}\label{eq:cov_binned}
        {\rm Cov}\left({\rm vec}\left(\tilde{\sf C}^{ab}_q\right)_A,{\rm vec}\left(\tilde{\sf C}^{fg}_{q'}\right)_F\right)=\sum_{{\bf l}\in q} B^{\bf l}_q\sum_{{\bf l}'\in q'} B^{{\bf l}'}_{q'}\,\Sigma^{AF}_{{\bf l}{\bf l}'}.
      \end{equation}
      which can then be used to estimate the covariance of the mode-decoupled bandpowers multiplying it by the inverse mode-coupling matrix twice. I.e., schematically:
      \begin{equation}
        {\rm Cov}\left(\hat{\sf C}\right)=\mathcal{M}^{-1}\cdot{\rm Cov}\left(\tilde{C}\right)\cdot\left(\mathcal{M}^{-1}\right)^{T},
      \end{equation}
      where we have suppressed all indices for simplicity. The problem of estimating the pseudo-$C_\ell$ covariance therefore reduces to estimating $\Sigma^{AC}_{{\bf l}{\bf l}'}$.
    
      We now make use of Wick's theorem, which states that, for Gaussian fields, $\langle a\,b\,f\,g\rangle=\langle a\,b\rangle\langle f\,g\rangle+\langle a\,f\rangle\langle b\,g\rangle+\langle a\,g\rangle\langle b\,f\rangle$. In this case, the expression for $\Sigma^{AF}_{{\bf l}{\bf l'}}$ reads:
      \begin{align}\nonumber
        \Sigma^{AF}_{{\bf l}{\bf l}'}=&\left[K\summ{k}\,^aM^{\alpha\alpha'}_{{\bf l}{\bf k}}\,^gM^{\gamma\gamma'*}_{{\bf l}'{\bf k}}C^{ag,(\alpha'\gamma')}_k\right]\left[K\summ{q}\,^bM^{\beta\beta'*}_{{\bf l}{\bf q}}\,^fM^{\phi\phi'}_{{\bf l}'{\bf q}}C^{bf,(\beta'\phi')}_q\right]^*\\\label{eq:covar_general}
        &+\left((g,\gamma)\leftrightarrow(f,\phi)\right),
      \end{align}
      where we implicitly sum over repeated indices (e.g. $\alpha'$), and the second term is equivalent to the first one after swapping the roles of fields ${\bf f}$ and ${\bf g}$. Without any further approximations, for each pair $({\bf l},{\bf l}')$, we would need to perform two 2-dimensional integrals, and therefore the calculation would scale like $\ell_{\rm max}^6$, quickly becoming unfeasible.
    
      Under the assumption that the coupling coefficients $M_{{\bf l}{\bf k}}$ are sharply peaked around ${\bf l}={\bf k}$, we can simplify the expression above approximating the power spectra as constants within the support of the coupling coefficients \cite{2004MNRAS.349..603E}. Explicitly, we approximate 
      \begin{equation}\nonumber
        C^{ag,(\alpha'\gamma')}_kC^{bf,(\beta',\phi')}_q\simeq C^{ag,(\alpha'\gamma')}_{(\ell}C^{bf,(\beta',\phi')}_{\ell')}\equiv\frac{1}{2}\left(C^{ag,(\alpha'\gamma')}_\ell C^{bf,(\beta',\phi')}_{\ell'}+C^{ag,(\alpha'\gamma')}_{\ell'} C^{bf,(\beta',\phi')}_\ell\right).
      \end{equation}
      In this case, the expression for $\Sigma^{AF}_{{\bf l}{\bf l}'}$ simplifies to
      \begin{align}\label{eq:covar_efstathiou}
        \Sigma^{AF}_{{\bf l}{\bf l}'}=K^2C^{ag,(\alpha'\gamma')}_{(\ell}C^{bf,(\beta',\phi')}_{\ell')}\,^{ag}W^{\alpha\gamma,\alpha'\gamma'}_{{\bf l}{\bf l}'}\left(\,^{bf}W^{\beta\phi,\beta'\phi'}_{{\bf l}{\bf l}'}\right)^*+\left((g,\gamma)\leftrightarrow(f,\phi)\right),
      \end{align}
      where we have defined the covariance coupling coefficients
      \begin{equation}\label{eq:covar_coupling}
        \,^{ab}W^{\alpha\beta,\alpha'\beta'}_{{\bf l}{\bf l}'}=\summ{k}\,^aM^{\alpha\alpha'}_{{\bf l}{\bf k}}\left(\,^bM^{\beta\beta'}_{{\bf l}'{\bf k}}\right)^*.
      \end{equation}

      In order to compute these coefficients, let us start by defining the quantities
      \begin{equation}
        \,^{ab}I^{\pm s_a,\pm s_b}_{{\bf l}{\bf l'}}\equiv\summ{k}\summ{x}\summ{y}w_a({\bf x})w_b({\bf y})\,q^{\pm s_a}_{{\bf l}{\bf k}}({\bf x})\,\left[q^{\pm s_b}_{{\bf l}'{\bf k}}({\bf y})\right]^*,
      \end{equation}
      where $s_a$ is the spin of field ${\bf a}$, and
      \begin{equation}
        q^{\pm s}_{{\bf l}{\bf k}}({\bf x})\equiv \frac{\beta_{\ell,s}\beta_{k,s}}{2}\left(\left[\eth^s q({\bf l},{\bf x})\right]^* \eth^sq({\bf k},{\bf x})\pm\left[\bar{\eth}^sq({\bf l},{\bf x})\right]^*\bar{\eth}^sq({\bf k},{\bf x})\right).
      \end{equation}

      Now, in what follows, we will be concerned with the auto- and
      cross-correlations of spin-0 and spin-2 fields. Thus, to simplify the
      notation, we will enumerate the different types of coupling coefficients
      that exist for a spin-0 field with a single component that we will call
      $\delta$, in analogy to the projected galaxy overensity, and for a
      spin-2 field, $\boldsymbol{\gamma}$, with $E$ and $B$ components, $\gamma_E$ and $\gamma_B$, in analogy to the cosmic shear field. With this setup, all the possible non-zero $W^{\alpha\beta,\alpha'\beta'}_{{\bf l}{\bf l}'}$ can be expressed in terms of the $I^{\pm s_a,\pm s_b}_{{\bf l}{\bf l}'}$ as follows:
      \begin{align}\label{eq:coeff}
        &W^{\delta\delta,\delta\delta}_{{\bf l},{\bf l}'}=I^{0,0}_{{\bf l}{\bf l}'}; \\
        &W^{\delta\gamma_E,\delta\gamma_E}_{{\bf l},{\bf l}'}=W^{\delta\gamma_B,\delta\gamma_B}_{{\bf l},{\bf l}'}=I^{0,+2}_{{\bf l},{\bf l}'};\hspace{12pt}
        W^{\delta\gamma_E,\delta\gamma_B}_{{\bf l},{\bf l}'}=-W^{\delta\gamma_B,\delta\gamma_E}_{{\bf l},{\bf l}'}=-i\,I^{0,-2}_{{\bf l},{\bf l}'};\\
        &W^{\gamma_E\gamma_E,\gamma_E\gamma_E}_{{\bf l},{\bf l}'}=W^{\gamma_E\gamma_B,\gamma_E\gamma_B}_{{\bf l},{\bf l}'}=W^{\gamma_B\gamma_E,\gamma_B\gamma_E}_{{\bf l},{\bf l}'}=W^{\gamma_B\gamma_B,\gamma_B\gamma_B}_{{\bf l},{\bf l}'}=I^{+2,+2}_{{\bf l}{\bf l}'}\\
        &W^{\gamma_E\gamma_E,\gamma_B\gamma_B}_{{\bf l},{\bf l}'}=W^{\gamma_B\gamma_B,\gamma_E\gamma_E}_{{\bf l},{\bf l}'}=-W^{\gamma_B\gamma_E,\gamma_E\gamma_B}_{{\bf l},{\bf l}'}=-W^{\gamma_E\gamma_B,\gamma_B\gamma_E}_{{\bf l},{\bf l}'}=I^{-2,-2}_{{\bf l}{\bf l}'}\\
        &W^{\gamma_E\gamma_E,\gamma_E\gamma_B}_{{\bf l},{\bf l}'}=-W^{\gamma_E\gamma_B,\gamma_E\gamma_E}_{{\bf l},{\bf l}'}=W^{\gamma_B\gamma_E,\gamma_B\gamma_B}_{{\bf l},{\bf l}'}=-W^{\gamma_B\gamma_B,\gamma_B\gamma_E}_{{\bf l},{\bf l}'}=-i\,I^{+2,-2}_{{\bf l}{\bf l}'}\\
        &W^{\gamma_E\gamma_E,\gamma_B\gamma_E}_{{\bf l},{\bf l}'}=W^{\gamma_E\gamma_B,\gamma_B\gamma_B}_{{\bf l},{\bf l}'}=-W^{\gamma_B\gamma_E,\gamma_E\gamma_E}_{{\bf l},{\bf l}'}=-W^{\gamma_B\gamma_B,\gamma_E\gamma_B}_{{\bf l},{\bf l}'}=i\,I^{-2,+2}_{{\bf l}{\bf l}'}.
      \end{align}
      Thus, in principle, we only need to compute 7 different types of terms
      ($I^{0,0}$, $I^{0,\pm2}$, $I^{\pm2,\pm2}$ and $I^{\pm2,\mp2})$). In
      order to simplify these expressions further, we follow \cite{Efstathiou:2006eb,2017A&A...602A..41C} and neglect all gradients of the masks\footnote{It is worth noting that it is possible in principle to avoid this approximation, as demonstrated in \cite{2005MNRAS.360..509C}.}, which allows us to relate the different $I^{\pm s_a,\pm s_b}$ through the following set of identities:
      \begin{align}\nonumber
        \summ{x}w({\bf x})\left(\bar{\eth}^2q({\bf l},{\bf x})\right)^*\,\bar{\eth}^2q({\bf k},{\bf x})
        &=\summ{x}w({\bf x})\eth^2 q^*({\bf l},{\bf x})\,\bar{\eth}^2q({\bf k},{\bf x})\\\nonumber
        &=\summ{x}q^*({\bf l},{\bf x})\,\eth^2\left(\bar{\eth}^2q({\bf k},{\bf x})w({\bf x})\right)\\\nonumber
        &\simeq\summ{x}q^*({\bf l},{\bf x})\,\left(\eth^2\bar{\eth}^2q({\bf k},{\bf x})\right)w({\bf x})\\
        &=\frac{1}{\beta^2_{k,2}}\summ{x}q^*({\bf l},{\bf x})\,q({\bf k},{\bf x})w({\bf x})\\\nonumber
        &=\summ{x}q^*({\bf l},{\bf x})\,\left(\bar{\eth}^2\eth^2q({\bf k},{\bf x})\right)w({\bf x})\\\nonumber
        &=\summ{x}\bar{\eth}^2\left(w({\bf x})\,q^*({\bf l},{\bf x})\right)\,\eth^2q({\bf k},{\bf x})\\
        &\simeq\summ{x}w({\bf x})\,\left(\eth^2\,q({\bf l},{\bf x})\right)^*\,\eth^2q({\bf k},{\bf x}),
      \end{align}
      where we have made repeated use of integration by parts.
    
      Using these identities together with the completeness relation of the basis functions, it is possible to simplify the expressions for the $I^{\pm s_a,\pm s_b}$:
      \begin{align}
        &\,^{ab}I^{0,0}_{{\bf l}{\bf l}'}=\,^{ab}I^{0,+2}_{{\bf l}{\bf l}'}=\,^{ab}J^0_{{\bf l}{\bf l}'}; \hspace{12pt}
        \,^{ab}I^{+2,+2}_{{\bf l}{\bf l}'}=\,^{ab}J^+_{{\bf l}{\bf l}'};\\
        &\,^{ab}I^{+2,-2}_{{\bf l}{\bf l}'}=\,^{ab}I^{-2,+2}_{{\bf l}{\bf l}'}=\,^{ab}J^-_{{\bf l}{\bf l}'}; \hspace{12pt}
        \,^{ab}I^{0,-2}_{{\bf l}{\bf l}'}=\,^{ab}I^{-2,-2}_{{\bf l}{\bf l}'}=0,
      \end{align}
      where  we have defined
      \begin{align}\label{eq:covar_coupling_J}
        &\,^{ab}J^0_{{\bf l}{\bf l}'}=\summ{x}\left(w_a w_b\right)({\bf x})\,q({\bf l},{\bf x})\,\left[q({\bf l}',{\bf x})\right]^*,\\
        &\,^{ab}J^\pm_{{\bf l}{\bf l}'}=\summ{x}\left(w_a w_b\right)({\bf x})\,q^{\pm2}_{{\bf l}{\bf l}'}({\bf x}).
      \end{align}
      Thus, the only surviving non-zero coupling coefficients are
      \begin{align}
        &W^{\delta\delta,\delta\delta}_{{\bf l},{\bf l}'}=W^{\delta\gamma_E,\delta\gamma_E}_{{\bf l},{\bf l}'}=W^{\delta\gamma_B,\delta\gamma_B}_{{\bf l},{\bf l}'}=J^0_{{\bf l}{\bf l}'};\\
        &W^{\gamma_E\gamma_E,\gamma_E\gamma_E}_{{\bf l},{\bf l}'}=W^{\gamma_E\gamma_B,\gamma_E\gamma_B}_{{\bf l},{\bf l}'}=W^{\gamma_B\gamma_E,\gamma_B\gamma_E}_{{\bf l},{\bf l}'}=W^{\gamma_B\gamma_B,\gamma_B\gamma_B}_{{\bf l},{\bf l}'}=J^+_{{\bf l}{\bf l}'}\\\nonumber
        &W^{\gamma_E\gamma_E,\gamma_E\gamma_B}_{{\bf l},{\bf l}'}=-W^{\gamma_E\gamma_B,\gamma_E\gamma_E}_{{\bf l},{\bf l}'}=W^{\gamma_B\gamma_E,\gamma_B\gamma_B}_{{\bf l},{\bf l}'}=-W^{\gamma_B\gamma_B,\gamma_B\gamma_E}_{{\bf l},{\bf l}'}=\\
        &\hspace{20pt}=-W^{\gamma_E\gamma_E,\gamma_B\gamma_E}_{{\bf l},{\bf l}'}=-W^{\gamma_E\gamma_B,\gamma_B\gamma_B}_{{\bf l},{\bf l}'}=W^{\gamma_B\gamma_E,\gamma_E\gamma_E}_{{\bf l},{\bf l}'}=W^{\gamma_B\gamma_B,\gamma_E\gamma_B}_{{\bf l},{\bf l}'}=-i\,J^-_{{\bf l}{\bf l}'}.
      \end{align}
    
      We have reduced the problem of computing the covariance in Eq. \ref{eq:covar_general} to the problem of computing the coupling coefficients \ref{eq:covar_coupling} entering Eq. \ref{eq:covar_efstathiou}, and we have now shown that there are only 3 independent coefficients, given by Eq. \ref{eq:covar_coupling_J}. In order to simplify the calculation further, it is now useful to inspect these results for the specific case of fields defined on the sphere.
    
      \subsubsection{Covariances for curved skies}
        As described in Section \ref{ssec:theory.pcl}, in the curved-sky
        estimator it is common to first average over the harmonic number $m$
        as part of the bandpower binning operation. Let us, therefore, define
        \begin{equation}
          \tilde{C}^{a_\alpha b_\beta}_\ell=\frac{1}{2\ell+1}\sum_{m=-\ell}^\ell (\tilde{a}_\alpha)_{\ell m} (\tilde{b}_\beta)_{\ell m}^*.
        \end{equation}
        The covariance of these objects, under the approximation in Eq. \ref{eq:covar_efstathiou}, can be computed as:
        \begin{equation}\label{eq:cov_sph}
          {\rm Cov}\left(\tilde{C}^{a_\alpha b_\beta}_\ell,\tilde{C}^{f_\phi g_\gamma}_{\ell'}\right) = C^{a_{\alpha'} f_{\phi'}}_{(\ell}C^{b_{\beta'} g_{\gamma'}}_{\ell')}\,^{af}_{bg}\mathcal{W}^{\alpha\phi,\alpha'\phi'}_{\beta\gamma,\beta'\gamma'}(\ell,\ell')+\left((g,\gamma)\leftrightarrow(f,\phi)\right),
        \end{equation}
        where we have defined the symbols
        \begin{equation}
          ^{af}_{bg}\mathcal{W}^{\alpha\phi,\alpha'\phi'}_{\beta\gamma,\beta'\gamma'}(\ell,\ell')\equiv\sum_{m=-\ell}^\ell \sum_{m'=-\ell'}^{\ell'}\frac{\,^{af}W^{\alpha\phi,\alpha'\phi'}_{{\bf l}{\bf l}'}\left(\,^{bg}W^{\beta\gamma,\beta'\gamma'}_{{\bf l}{\bf l}'}\right)^*}{(2\ell+1)(2\ell'+1)}.
        \end{equation}
        These quantities involve terms of the form:
        \begin{equation}
          \,^{af}_{bg}\mathcal{J}^{XY}_{\ell\ell'}\equiv\sum_{m=-\ell}^\ell \sum_{m'=-\ell'}^{\ell'}\frac{\,^{af}J^X_{{\bf l}{\bf l}'}\left(\,^{bg}J^Y_{{\bf l}{\bf l}'}\right)^*}{(2\ell+1)(2\ell'+1)}
        \end{equation}
        Making use of the completeness relation for the Wigner 3-j symbols, it is possible to show, that these can be written as:
        \begin{align}\label{eq:coeff_cmcm}
          &\,^{af}_{bg}\mathcal{J}^{00}_{\ell\ell'}=\Xi^{00}_{\ell\ell'}(w_aw_f,w_bw_g),\hspace{12pt}
           \,^{af}_{bg}\mathcal{J}^{0+}_{\ell\ell'}=\Xi^{0+}_{\ell\ell'}(w_aw_f,w_bw_g),\\
          &\,^{af}_{bg}\mathcal{J}^{++}_{\ell\ell'}=\Xi^{++}_{\ell\ell'}(w_aw_f,w_bw_g),\hspace{12pt}
           \,^{af}_{bg}\mathcal{J}^{--}_{\ell\ell'}=\Xi^{--}_{\ell\ell'}(w_aw_f,w_bw_g),\\
          &\,^{af}_{bg}\mathcal{J}^{+-}_{\ell\ell'}=
           \,^{af}_{bg}\mathcal{J}^{-+}_{\ell\ell'}=
           \,^{af}_{bg}\mathcal{J}^{0-}_{\ell\ell'}=0,
        \end{align}
        where the $\Xi^{XY}_{\ell\ell'}$ are defined in Eq. \ref{eq:coeff_mcm}.
        Thus, computationally speaking, the problem of computing covariance matrices reduces to that of computing the same coupling coefficients needed for the computation of the pseudo-$C_\ell$ power spectra themselves, except now they involve product of two masks, rather than the masks alone. Using the same notation as in Eq. \ref{eq:coeff}, the only non-zero $\mathcal{W}$ coefficients are:
        \begin{align}
          &\mathcal{W}^{\delta  \delta  ,\delta  \delta  }_{\delta  \delta  ,\delta  \delta  }=
           \mathcal{W}^{\delta  \delta  ,\delta  \delta  }_{\delta  \gamma_X,\delta  \gamma_X}=
           \mathcal{W}^{\delta  \gamma_X,\delta  \gamma_X}_{\delta  \gamma_Y,\delta  \gamma_Y}=\mathcal{J}^{00}\\
          &\mathcal{W}^{\delta  \delta  ,\delta  \delta  }_{\gamma_X\gamma_Y,\gamma_X\gamma_Y}=
           \mathcal{W}^{\delta  \gamma_Z,\delta  \gamma_Z}_{\gamma_X\gamma_Y,\gamma_X\gamma_Y}=\mathcal{J}^{0+}\\
          &\mathcal{W}^{\gamma_X\gamma_Y,\gamma_X\gamma_Y}_{\gamma_W\gamma_Z,\gamma_W\gamma_Z}=\mathcal{J}^{++}\\\nonumber
          &\mathcal{W}^{\gamma_X\gamma_E,\gamma_X\gamma_B}_{\gamma_Y\gamma_E,\gamma_Y\gamma_B}=
           \mathcal{W}^{\gamma_X\gamma_E,\gamma_X\gamma_B}_{\gamma_B\gamma_Y,\gamma_E\gamma_Y}=
           \mathcal{W}^{\gamma_B\gamma_X,\gamma_E\gamma_X}_{\gamma_B\gamma_Y,\gamma_E\gamma_Y}=\\
          &\hspace{12pt}
           \mathcal{W}^{\gamma_X\gamma_B,\gamma_X\gamma_E}_{\gamma_Y\gamma_B,\gamma_Y\gamma_E}=
           \mathcal{W}^{\gamma_X\gamma_B,\gamma_X\gamma_E}_{\gamma_E\gamma_Y,\gamma_B\gamma_Y}=
           \mathcal{W}^{\gamma_E\gamma_X,\gamma_B\gamma_X}_{\gamma_E\gamma_Y,\gamma_B\gamma_Y}=-\mathcal{J}^{--}\\\nonumber
          &\mathcal{W}^{\gamma_X\gamma_B,\gamma_X\gamma_E}_{\gamma_Y\gamma_E,\gamma_Y\gamma_B}=
           \mathcal{W}^{\gamma_X\gamma_B,\gamma_X\gamma_E}_{\gamma_B\gamma_Y,\gamma_E\gamma_Y}=
           \mathcal{W}^{\gamma_E\gamma_X,\gamma_B\gamma_X}_{\gamma_B\gamma_Y,\gamma_E\gamma_Y}=\\
          &\hspace{12pt}
           \mathcal{W}^{\gamma_X\gamma_E,\gamma_X\gamma_B}_{\gamma_Y\gamma_B,\gamma_Y\gamma_E}=
           \mathcal{W}^{\gamma_X\gamma_E,\gamma_X\gamma_B}_{\gamma_E\gamma_Y,\gamma_B\gamma_Y}=
           \mathcal{W}^{\gamma_B\gamma_X,\gamma_E\gamma_X}_{\gamma_E\gamma_Y,\gamma_B\gamma_Y}=\mathcal{J}^{--},
        \end{align}
        where $(X,Y,Z,W)$ stand for either $E$ or $B$, and where we have suppressed all redundant indices (including $\ell\ell'$). Any pseudo-$C_\ell$ covariance element can then be found by replacing these results in Eq. \ref{eq:cov_sph}. Some explicit examples for common terms can be found in Appendix A.3.3 of \cite{2017A&A...602A..41C}.

      \subsubsection{Covariances for flat skies}
        Similar results hold in the case of flat skies. As mentioned in
        Section \ref{ssec:theory.pcl}, in this case averaging over the
        Fourier-space azimuth happens while binning into bandpowers. Under the
        assumption that the underlying power spectra are roughly constant
        within each bandpower, in this case, the covariance matrix of the bandpowers defined in Eq. \ref{eq:bandpowers} takes the form:
        \begin{equation}\label{eq:cov_flat}
          {\rm Cov}\left(\tilde{C}^{a_\alpha b_\beta}_q,\tilde{C}^{f_\phi g_\gamma}_{q'}\right) = C^{a_{\alpha'} f_{\phi'}}_{(q}C^{b_{\beta'} g_{\gamma'}}_{q')}\sum_{{\bf l}\in q} \frac{1}{N_q} \sum_{{\bf l}' \in q'} \frac{1}{N_{q'}}\,\,^{af}_{bg}\bar{\mathcal{W}}^{\alpha\phi,\alpha'\phi'}_{\beta\gamma,\beta'\gamma'}({\bf l},{\bf l}')+\left((g,\gamma)\leftrightarrow(f,\phi)\right),
        \end{equation}
        where the coefficients $\bar{\mathcal{W}}$ are related to the
        mode-coupling coefficients $\bar{\Xi}$ defined in
        Eq.~\ref{eq:coeff_mcm_flat}, in the same way that the curved-sky coefficients $\mathcal{W}$ were related to the $\Xi$:
        \begin{align}
          &\bar{\mathcal{W}}^{\delta  \delta  ,\delta  \delta  }_{\delta  \delta  ,\delta  \delta  }=
           \bar{\mathcal{W}}^{\delta  \delta  ,\delta  \delta  }_{\delta  \gamma_X,\delta  \gamma_X}=
           \bar{\mathcal{W}}^{\delta  \gamma_X,\delta  \gamma_X}_{\delta  \gamma_Y,\delta  \gamma_Y}=\bar{\mathcal{J}}^{00}\\
          &\bar{\mathcal{W}}^{\delta  \delta  ,\delta  \delta  }_{\gamma_X\gamma_Y,\gamma_X\gamma_Y}=
           \bar{\mathcal{W}}^{\delta  \gamma_Z,\delta  \gamma_Z}_{\gamma_X\gamma_Y,\gamma_X\gamma_Y}=\bar{\mathcal{J}}^{0+}\\
          &\bar{\mathcal{W}}^{\gamma_X\gamma_Y,\gamma_X\gamma_Y}_{\gamma_W\gamma_Z,\gamma_W\gamma_Z}=\bar{\mathcal{J}}^{++}\\\nonumber
          &\bar{\mathcal{W}}^{\gamma_X\gamma_E,\gamma_X\gamma_B}_{\gamma_Y\gamma_E,\gamma_Y\gamma_B}=
           \bar{\mathcal{W}}^{\gamma_X\gamma_E,\gamma_X\gamma_B}_{\gamma_B\gamma_Y,\gamma_E\gamma_Y}=
           \bar{\mathcal{W}}^{\gamma_B\gamma_X,\gamma_E\gamma_X}_{\gamma_B\gamma_Y,\gamma_E\gamma_Y}=\\
          &\hspace{12pt}
           \bar{\mathcal{W}}^{\gamma_X\gamma_B,\gamma_X\gamma_E}_{\gamma_Y\gamma_B,\gamma_Y\gamma_E}=
           \bar{\mathcal{W}}^{\gamma_X\gamma_B,\gamma_X\gamma_E}_{\gamma_E\gamma_Y,\gamma_B\gamma_Y}=
           \bar{\mathcal{W}}^{\gamma_E\gamma_X,\gamma_B\gamma_X}_{\gamma_E\gamma_Y,\gamma_B\gamma_Y}=-\bar{\mathcal{J}}^{--}\\\nonumber
          &\bar{\mathcal{W}}^{\gamma_X\gamma_B,\gamma_X\gamma_E}_{\gamma_Y\gamma_E,\gamma_Y\gamma_B}=
           \bar{\mathcal{W}}^{\gamma_X\gamma_B,\gamma_X\gamma_E}_{\gamma_B\gamma_Y,\gamma_E\gamma_Y}=
           \bar{\mathcal{W}}^{\gamma_E\gamma_X,\gamma_B\gamma_X}_{\gamma_B\gamma_Y,\gamma_E\gamma_Y}=\\
          &\hspace{12pt}
           \bar{\mathcal{W}}^{\gamma_X\gamma_E,\gamma_X\gamma_B}_{\gamma_Y\gamma_B,\gamma_Y\gamma_E}=
           \bar{\mathcal{W}}^{\gamma_X\gamma_E,\gamma_X\gamma_B}_{\gamma_E\gamma_Y,\gamma_B\gamma_Y}=
           \bar{\mathcal{W}}^{\gamma_B\gamma_X,\gamma_E\gamma_X}_{\gamma_E\gamma_Y,\gamma_B\gamma_Y}=\bar{\mathcal{J}}^{--},
        \end{align}
        and, as before,
        \begin{align}\label{eq:coeff_cmcm_flat}
          &\,^{af}_{bg}\bar{\mathcal{J}}^{00}_{{\bf l}{\bf l}'}=\bar{\Xi}^{00}_{{\bf l}{\bf l}'}(w_aw_f,w_bw_g),\hspace{12pt}
           \,^{af}_{bg}\bar{\mathcal{J}}^{0+}_{{\bf l}{\bf l}'}=\bar{\Xi}^{0+}_{{\bf l}{\bf l}'}(w_aw_f,w_bw_g),\\
          &\,^{af}_{bg}\bar{\mathcal{J}}^{++}_{{\bf l}{\bf l}'}=\bar{\Xi}^{++}_{{\bf l}{\bf l}'}(w_aw_f,w_bw_g),\hspace{12pt}
           \,^{af}_{bg}\bar{\mathcal{J}}^{--}_{{\bf l}{\bf l}'}=\bar{\Xi}^{--}_{{\bf l}{\bf l}'}(w_aw_f,w_bw_g),\\
          &\,^{af}_{bg}\bar{\mathcal{J}}^{+-}_{{\bf l}{\bf l}'}=
           \,^{af}_{bg}\bar{\mathcal{J}}^{-+}_{{\bf l}{\bf l}'}=
           \,^{af}_{bg}\bar{\mathcal{J}}^{0-}_{{\bf l}{\bf l}'}=0.
        \end{align}
        
    \subsection{Approximate covariances}\label{ssec:theory.approx}
      When presenting our results in Section \ref{sec:results}, we will compare the true covariance matrix, estimated from a large number of Gaussian simulations, with the analytical covariance estimated under different approximations. In descending order of complexity, these are:
      \begin{enumerate}
        \item The narrow-kernel approximation (labeled {\bf NKA} here), described in the previous sections. This approximation assumes that the support of the harmonic-space masks (represented by the mode-coupling coefficients in e.g. Eq. \ref{eq:mcms}) is small compared to the variation of the true power spectrum with $\ell$. Additionally, it neglects all derivatives of the sky mask when accounting for the spin nature of the fields involved.
        \item The {\bf spin-0} approximation corresponds to a simplified version of the NKA in which the spin nature of all fields involved is completely ignored, and all fields, including the $E$- and $B$-mode components of a spin-2 field, are treated as spin-0 quantities.
        \item The mode-counting approximation (labeled {\bf MC} here),
          commonly known as the {\sl Knox formula} \cite{1995PhRvD..52.4307K},
          which applies the result found for full-sky observations (Eq.
          \ref{eq:cov_naive}) to masked fields, corrected by an overall factor that accounts to the loss of modes due to the sky mask. In this case, the covariance is simply given by:
        \begin{equation}\label{eq:cov_naive_bpw}
          {\rm Cov}\left(C^{ab}_q,C^{cd}_{q'}\right)=\delta^K_{qq'}\frac{C^{ad}_qC^{bc}_q+C^{ac}_qC^{bd}_q}{(2\ell_q+1)\,f_{\rm sky}\,N_q},
        \end{equation}
        where $\ell_q$ is the mean multipole in the $q$-th bandpower, $N_q$ is the number of multipoles assigned to it, and $f_{\rm sky}$ is the available sky fraction.
      \end{enumerate}

  \section{Results}\label{sec:results}
    \subsection{Simulations}\label{ssec:results.sims}
      In order to quantitatively study the performance of the analytical approximations for the power spectrum covariance matrix introduced in Section \ref{sec:theory}, we generate a large number of Gaussian simulations including a number of realistic observational effects.
      
      Each simulation is a set of maps corresponding to a number of spin-0 and spin-2 fields that are drawn as Gaussian random fields following a set of input power spectra that include all relevant cross-correlations between different fields. We generate simulations for two types of fields modeled after the two main large-scale structure observables of photometric redshift surveys:
      \begin{itemize}
        \item Spin-0 fields, corresponding to maps of the overdensity of galaxies within a given redshift bin projected on the sphere. In keeping with the notation introduced in Section \ref{ssec:theory.pclcov}, we will label these fields as $\delta^a$, where the index $a$ denotes the redshift bin.
        \item Spin-2 fields, corresponding to maps of the cosmic shear measured from the projected shapes of galaxies in a given redshift bin. We will label these fields $\boldsymbol{\gamma}^a=(\gamma_E^a,\gamma_B^a)$ where, again, the index $a$ denotes the redshift bin.
      \end{itemize}
      The cross-correlation between two of these fields can be written as:
      \begin{equation}
        C^{ab}_\ell = S^{ab}_\ell+N^{ab}_\ell,
      \end{equation}
      where $S$ and $N$ are the power spectra of the cosmological signal and noise respectively. We model the signal part for $\delta$ and $\gamma_E$ as:
      \begin{equation}
        S^{ab}_\ell = \int d\chi \frac{W^a_\ell(\chi)\,W^b_\ell(\chi)}{\chi^2}\,P\left(k=\frac{\ell+1/2}{\chi},z\right),
      \end{equation}
      where $\chi$ is the comoving radial distance, $z\equiv z(\chi)$ is the corresponding redshift in the lightcone, $P(k,z)$ is the matter power spectrum. The window functions are given by \cite{2004PhRvD..70d3009H}\footnote{See \cite{2019ApJS..242....2C} and references therein for details about these calculations.}:
      \begin{align}
        \nonumber
        &W^\delta_\ell(\chi)=b(z)\,H(z)\,p_z(z),\\
        &W^\gamma_\ell(\chi)=f_\ell\,\frac{3H_0^2\Omega_M}{2}(1+z)\chi\,\int dz'\,p_z(z')\,\frac{\chi(z')-\chi}{\chi(z')},
      \end{align}
      where $H(z)$ is the expansion rate in units where the speed of light is $c=1$, $H_0\equiv H(z=0)$, $b(z)$ is the linear galaxy bias, $p_z$ is the normalized redshift distribution of galaxies within the redshift bin, and
      \begin{equation}
        f_\ell\equiv\frac{\sqrt{(\ell+2)(\ell+1)\ell(\ell-1)}}{(\ell+1/2)^2}.
      \end{equation}
      We assume zero signal for the shear $B$-modes.
      \begin{figure}
        \centering
        \includegraphics[width=0.6\textwidth]{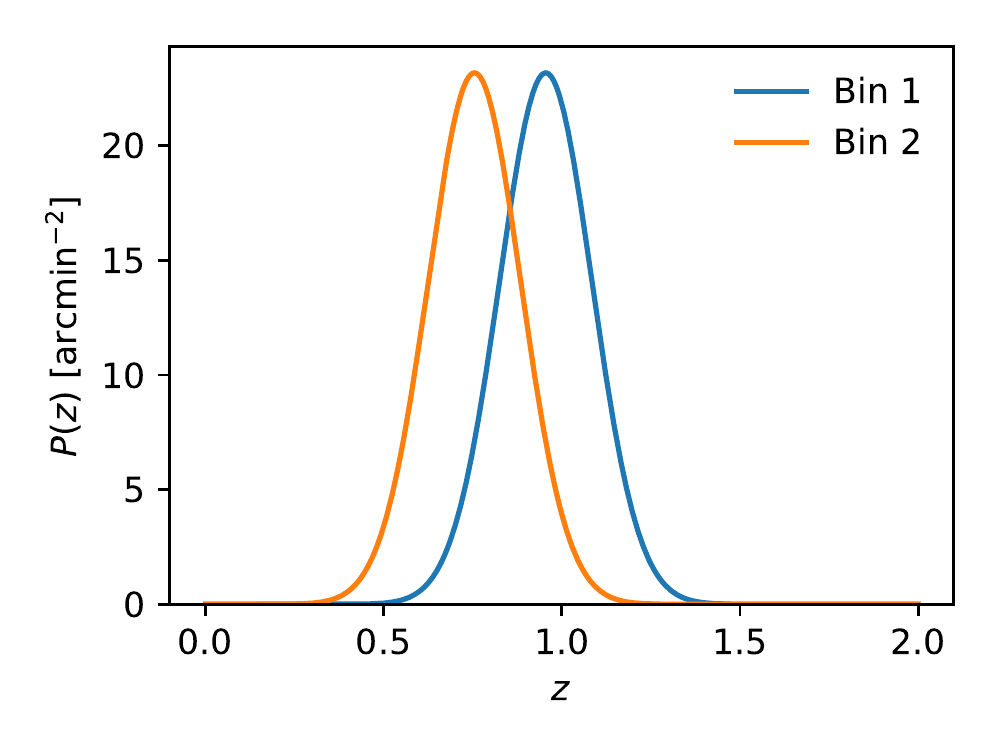}
        \caption{Redshift distributions assumed for the Gaussian simulations used in this analysis.}
        \label{fig:pz}
      \end{figure}
      \begin{figure*}
        \centering
        \includegraphics[width=\textwidth]{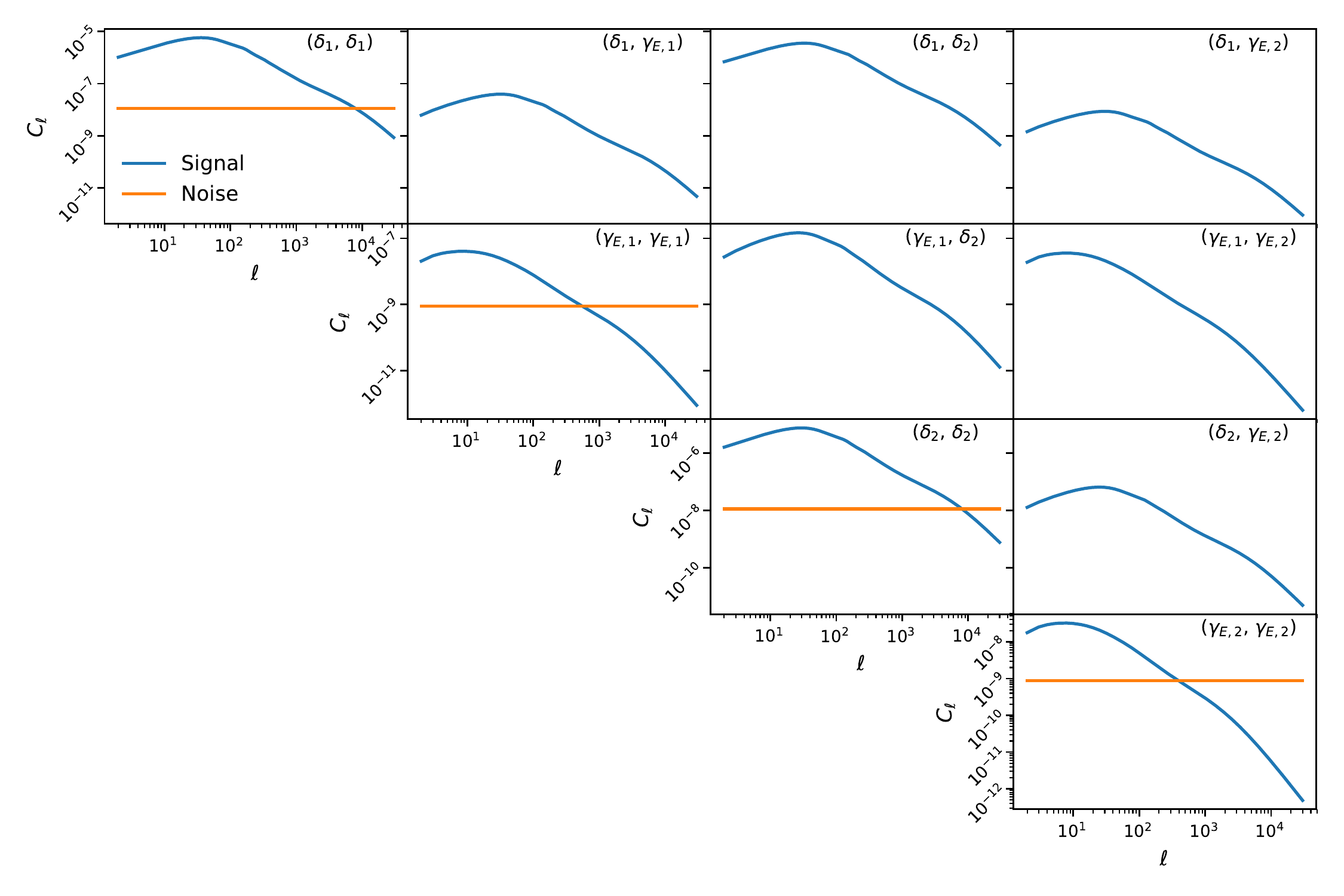}
        \caption{Signal (blue) and noise (orange) power spectra for the different observables used in our simulations. Note that the signal for shear $B$-modes ($\gamma_B$) is zero, and the noise is the same as that of $\gamma_E$. The different panels show different cross-correlations between $\delta$ and $\gamma_E$ in two redshift bins.}
        \label{fig:cl-2bins}
      \end{figure*}      
      \begin{figure}
        \centering
        \includegraphics[width=0.6\textwidth]{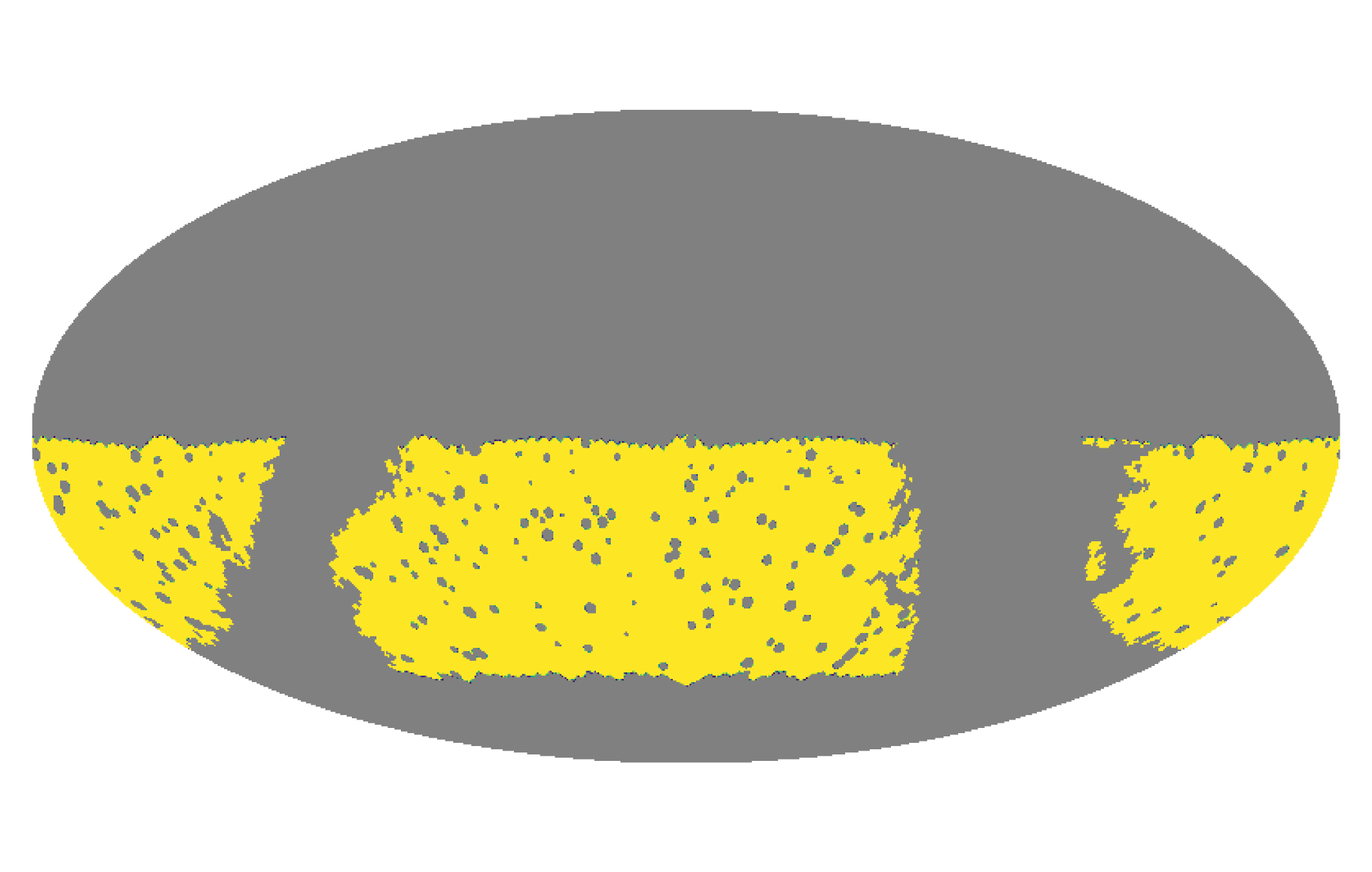}
        \includegraphics[width=0.6\textwidth]{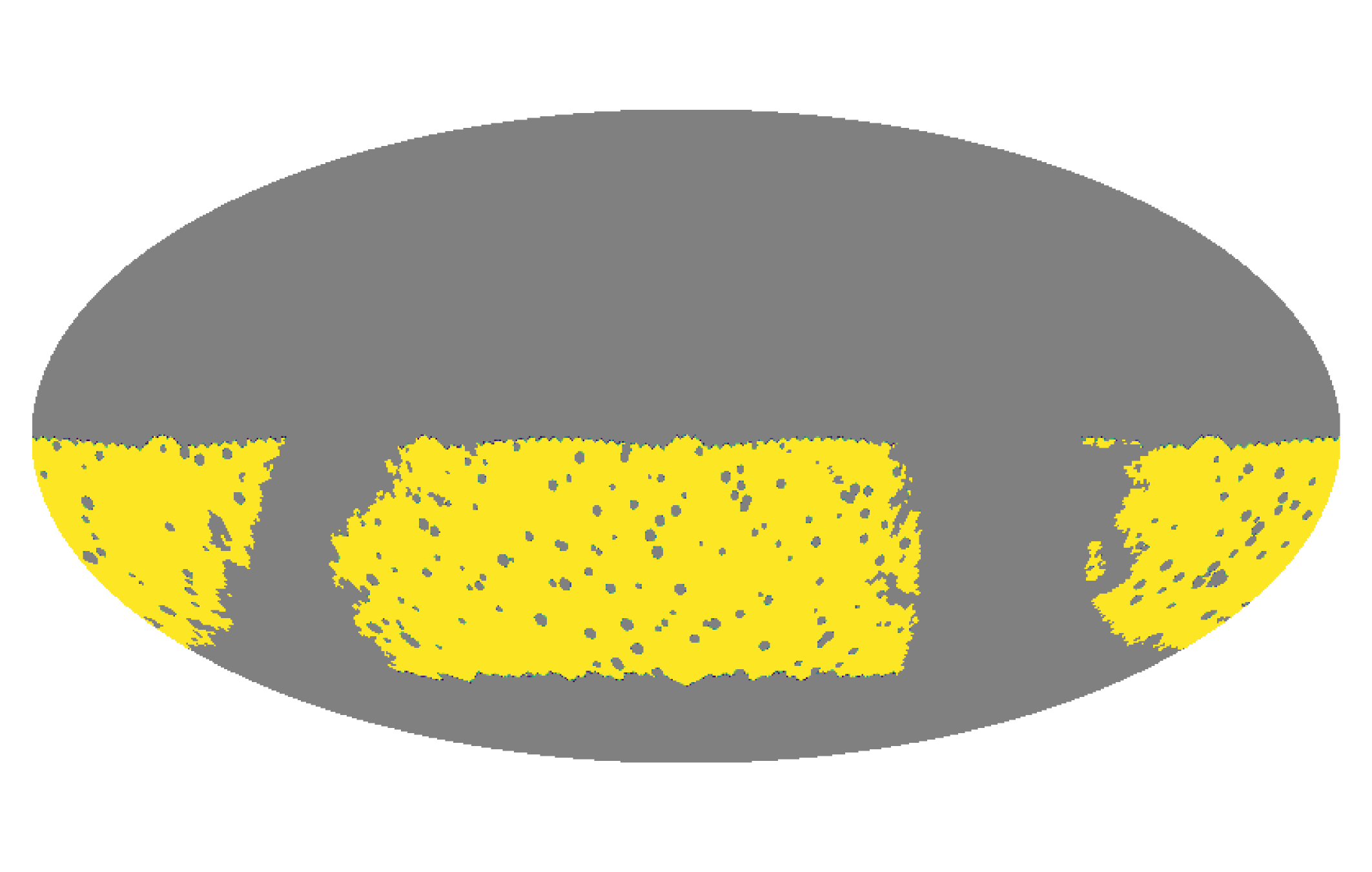}
        \caption{Sky masks used in our analysis for bins 1 and 2 (upper and lower panels respectively). The same mask is assumed for the galaxy overdensity $\delta$ and cosmic shear $\boldsymbol{\gamma}$ for simplicity.} \label{fig:mask}
      \end{figure}
      \begin{figure}
        \centering
        \includegraphics[width=0.7\textwidth]{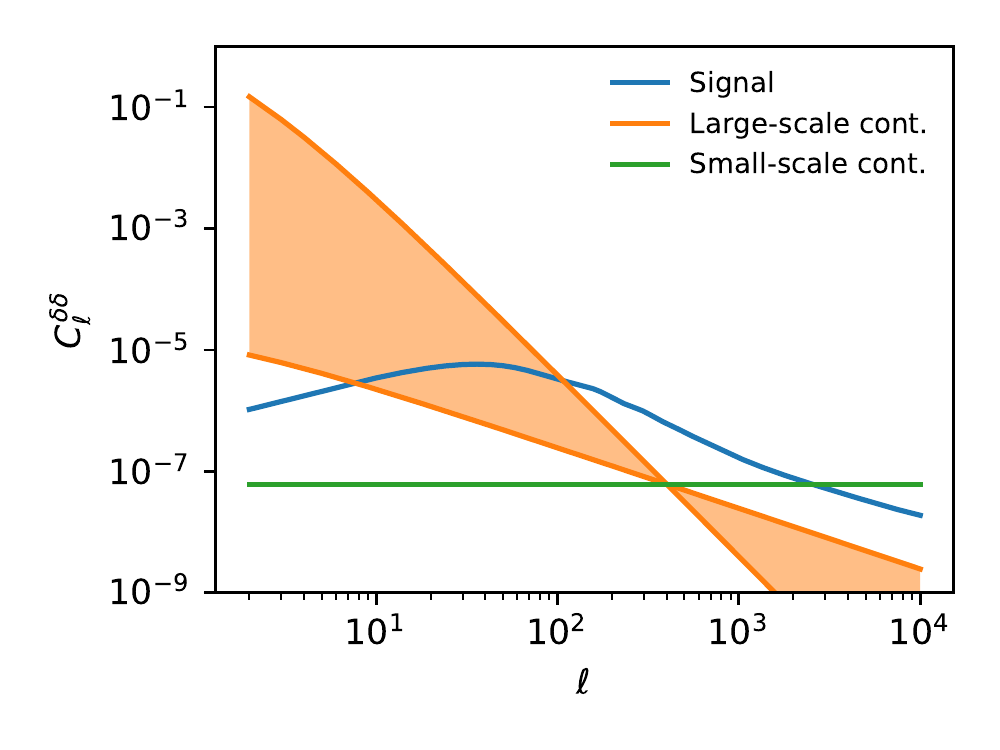}
        \caption{Input (signal $+$ noise) power spectrum for the galaxy overdensity in the first redshift bin (blue line). The orange band shows the range of power spectra used for the 100 large-scale contaminant templates used to study the impact of contaminant deprojection on the covariance matrix. The green line shows the power spectrum used to generate small-scale contaminants. The amplitudes of the contaminant power spectra were fixed by imposing a $10\%$ contamination level in the power spectrum at $\ell=400$. The same approach was used for the cosmic shear maps, with independent contaminants defined in $\gamma_E$ and $\gamma_B$.} \label{fig:clcont}
      \end{figure}

      The noise power spectrum is diagonal (i.e. zero between different fields and redshift bins), and is given by:
      \begin{equation}
        N_\ell^{\delta\delta} = \frac{1}{n_\Omega},\hspace{12pt}
        N_\ell^{\gamma_E\gamma_E} = N_\ell^{\gamma_B\gamma_B} = \frac{\sigma_\gamma^2}{n_\Omega},
      \end{equation}
      where $n_\Omega$ is the mean number density of galaxies in units of sterad$^{-2}$ (see below), and $\sigma_\gamma=0.28$ is the intrinsic shape scatter per ellipticity component.

      We consider the case of auto- and cross-correlations between all fields in two redshift bins, with redshift distributions modeled as Gaussians with width $\sigma_z=0.13$ centered around redshifts $0.75$ and $0.95$, with number densities $n_\Omega=7.5\,{\rm arcmin}^{-2}$. The corresponding redshift distributions $p_z(z)$ are shown in Figure \ref{fig:pz}. For simplicity we use a constant bias $b(z)=1$. For these specifications, we generate signal power spectra using the Core Cosmology Library ({\tt CCL} \cite{2019ApJS..242....2C}) for cosmological parameters $(\Omega_c,\Omega_b,h,A_s,n_s)=(0.27,0.045,0.67,2.1\times10^{-9},0.96)$. The resulting non-zero signal and noise power spectra are shown in Figure~\ref{fig:cl-2bins}.

      After generating a set of Gaussian maps, we mask them making use of a realistic sky mask. This mask has three main components: a cut in declination based on the expected sky coverage of LSST \cite{2014SPIE.9150E..15D}, a more conservative Galactic cut using the dust reddening data from \cite{1998ApJ...500..525S} and a set of 100 randomly positioned holes with a radius of 1 degree. To explore the case of cross-correlations between fields with different masks, we generated two masks for the two redshift bins above, consisting of two different sets of random holes. These masks are shown in Figure \ref{fig:mask}.

      We have also explored the impact of the presence of sky contaminants on the estimate of the covariance matrix. As described in \cite{2019MNRAS.484.4127A}, the presence of a small contamination from observational systematics in the data can be accounted for through a technique known as {\sl mode deprojection} \cite{1992ApJ...398..169R,2004PhRvD..69l3003S,2017MNRAS.465.1847E}. In this method, the contamination is modeled as a linear contribution at the map level from contaminants with a known template. Mode deprojection then consists on projecting the data onto the subspace of modes that are perpendicular to those templates, effectively removing all modes from the map that ``look like'' any of the contaminants. This removal of modes has been shown to provide unbiased estimates of the power spectrum, however it could potentially affect the power spectrum uncertainties due to the loss of statistical power. To study this effect, we have also generated contaminant maps of two types:
      \begin{itemize}
        \item {\bf Large-scale contaminants:} Gaussian random maps with a red spectrum of the form $C_\ell\propto (\ell+1)^\beta$, where $\beta$ is a random number chosen within the range $\beta\in(-1,-3)$.
        \item {\bf Small-scale contaminants:} Gaussian random maps with a flat spectrum $C_\ell={\rm const.}$
      \end{itemize}
      In both cases, we fixed the amplitude of the contaminant power spectrum such that it would yield a $10\%$ contamination in the data power spectrum at $\ell=400$. For spin-2 fields, we assumed the same power spectrum for $E$ and $B$ modes, with no cross-correlation between them. The resulting contaminant power spectra are shown in Figure \ref{fig:clcont}. When exploring the effects of mode deprojection, we generated 100 contaminant maps of both types and added them to each simulated realization. We then deprojected the full set of 100 contaminant templates from the simulated maps and computed the corresponding unbiased power spectra. Note that, in what follows, our fiducial results do not include the effects of mode deprojection. These are discussed separately.

      For our fiducial results, we generated a set of 20,000 random
      simulations. This number was chosen in order to recover the covariance
      matrix for all possible auto- and cross-correlations with sufficient
      accuracy. All simulations were generated as {\tt
        HEALPix}\footnote{\url{http://healpix.sourceforge.net}}~\cite{Gorski:2004by} maps with resolution $N_{\rm side}=512$. We then used {\tt NaMaster} to compute all possible power spectra for each simulation using narrow bandpowers of width $\Delta\ell=3$ from $\ell=2$ to $\ell=1023$. Finally, we used the power spectra from the simulations to estimate the sample covariance matrix
      \begin{equation}
        {\rm Cov}({\bf C}) =\frac{1}{N_{\rm sim}-1}\sum_{i=1}^{N_{\rm sim}}\left({\bf C}_i-\bar{\bf C}\right)\cdot\left({\bf C}_i-\bar{\bf C}\right)^T,
      \end{equation}
      where ${\bf C}_i$ is the vector of all possible power spectra for the $i$-th realization, and $\bar{\bf C}$ is the mean of this vector over all realizations. The comparison of this sample covariance with the analytical approximations described above is presented in the next sections.

    \subsection{Qualitative comparison}\label{ssec:results.visual}
      \begin{figure}
        \centering
        \includegraphics[width=\textwidth]{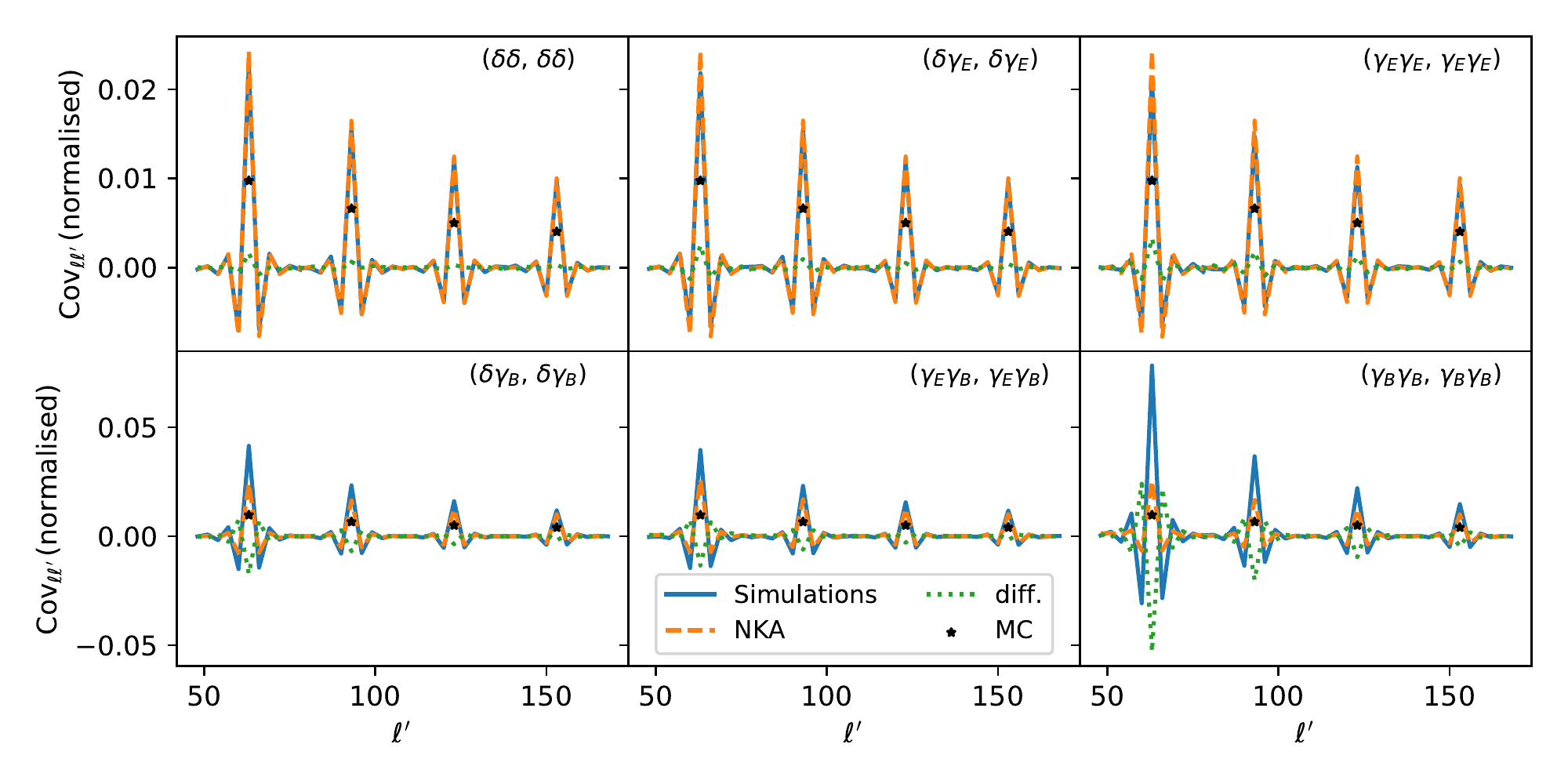}
        \caption{Four rows of the covariance matrix for different power spectra. The top panels show the cases of spectra with non-zero signal, involving $\delta$ and $\gamma_E$, while the bottom panels show cases involving $B$-modes. In each panel we show rows of the covariance matrix for $\ell=60,\,90,\,120$ and $150$, which peak at those central values. The different lines show the results for the sample covariance matrix (solid blue), its NKA estimator (dashed orange) and the difference between both (dotted green). The black stars show the mode-counting approximation to the covariance matrix (Eq. \ref{eq:cov_naive_bpw}). To facilitate the visualization of the different rows, we have divided them by $C_\ell^{aa}C_\ell^{bb}+(C^{ab}_\ell)^2$, the numerator of Eq.~\ref{eq:cov_naive_bpw}, for the two fields, $a$ and $b$, that are being correlated in each case. The NKA method is able to recover the covariance with high accuracy for all field combinations that do not involve $B$ modes, yielding visibly poorer results otherwise.}
        \label{fig:rows_1bin}
      \end{figure}
      \begin{figure}
        \centering
        \includegraphics[width=0.7\textwidth]{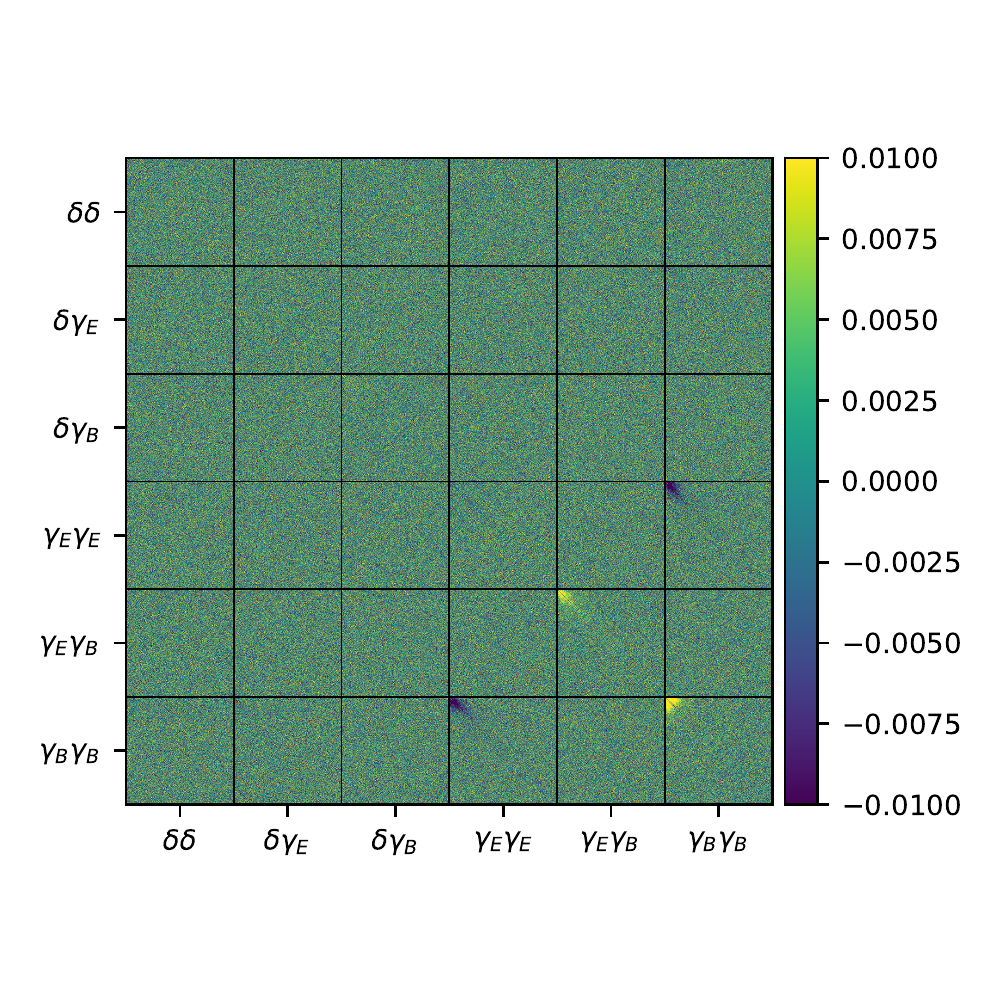}
        \caption{Difference between the correlation matrices associated to the sample covariance matrix and its estimate with the NKA method including all possible correlations between $\delta$ and $\boldsymbol{\gamma}$ in a single redshift bin. While the NKA estimator is able to recover the covariance matrix to high accuracy in most cases, it is not able to reproduce the off-diagonal correlations between different bandpowers in cases involving $B$ modes.} \label{fig:diff_corr_1bin}
      \end{figure}
      \begin{figure}
        \centering
        \includegraphics[width=0.9\textwidth]{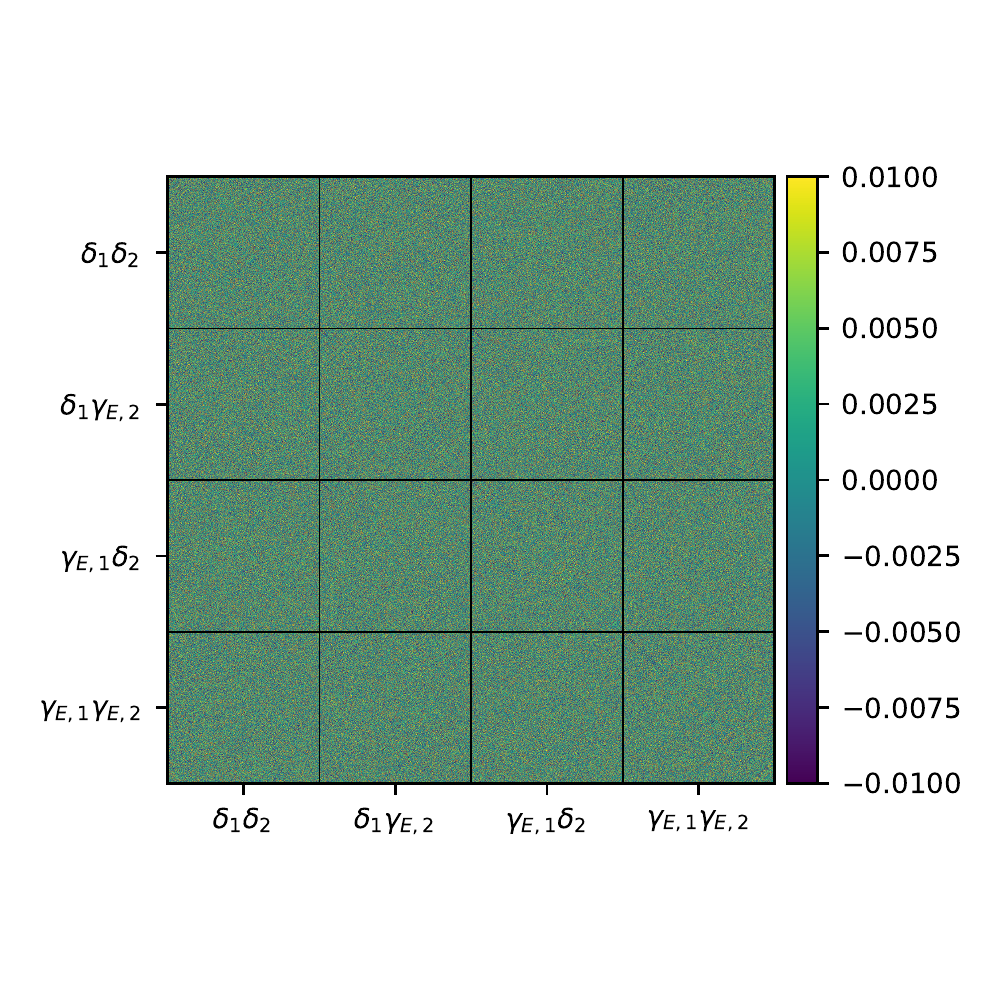}
        \caption{Same as Fig. \ref{fig:diff_corr_1bin} for all cross-correlations between $\delta$ and $\gamma_E$ measured in two different redshift bins.}
        \label{fig:corr_diff_2bins}
      \end{figure}
      As a first step, we visually compare the main properties of the sample covariance matrix estimated from the simulations and the NKA, Spin-0 and MC approximations described in the previous section. Figure \ref{fig:rows_1bin} shows four rows of the covariance matrix of different auto- and cross-correlations. The rows correspond to bandpowers centered on multipoles $\ell_q\simeq60,\,90,\,120$ and $150$. The upper panels show results for the non-zero power spectra ($\delta$-$\delta$, $\delta$-$\gamma_E$ and $\gamma_E$-$\gamma_E$), with the solid blue, dashed orange and dotted green lines showing results for the sample covariance matrix, the analytical covariance using the NKA approximation and their difference, respectively. For comparison, the black stars show the diagonal covariance matrix elements predicted by the MC approximation (Eq. \ref{eq:cov_naive_bpw}). We find an excellent agreement between the simulated and analytical covariances, with very small deviations in the amplitude of the diagonal and first few off-diagonal elements.
      
      The bottom panel in Figure~\ref{fig:rows_1bin} shows the same rows of the covariance matrix for power spectra involving $B$-modes (and therefore with zero signal expectation value). In this case we find significant differences, at the level of $30-50\%$, on the covariance matrix elements, with the analyticial prediction underestimating the error bars overall. This is expected and can be understood as follows: the presence of a sky mask mixes $E$ and $B$ modes. Although this mixing can be accounted for at the level of the power spectrum through the pseudo-$C_\ell$ estimator, the leaked modes contribute to the variance. This is particularly significant for power spectra involving $B$-modes, since the $E$-mode amplitude is significantly larger, especially at $\ell\lesssim200$, as can be seen in Fig. \ref{fig:cl-2bins}. Thus, if the effects of $E$-$B$ mixing caused by the sky mask are not accurately accounted for in the estimation of the covariance matrix for power spectra involving $B$-mode maps, we can expect a misestimation of the contribution to the covariance from the leaked $E$ modes that would underpredict the uncertainties. This is not a problem for power spectra involving only $E$ modes, since the only $B$ modes that leak into them are those associated with noise, and they have the same amplitude as the noise $E$ modes move into the $B$-mode map.

      This is further illustrated by Figure~\ref{fig:diff_corr_1bin}. The figure shows, for the case of a single redshift bin, the difference between the correlation matrices associated with the sample covariance matrix and the NKA estimate\footnote{The correlation matrix is defined as $r_{ij} = {\rm Cov}_{ij}/\sqrt{{\rm Cov}_{ii}{\rm Cov}_{jj}}$}. While the differences between both matrices are small for all elements involving $\delta$ and $\gamma_E$, all terms involving $B$-modes show a significant disagreement, particularly the  $\gamma_E\gamma_E$-$\gamma_B\gamma_B$, $\gamma_E\gamma_B$-$\gamma_E\gamma_B$ and $\gamma_B\gamma_B$-$\gamma_B\gamma_B$ boxes.

      We thus conclude that while the NKA estimator is able to recover the covariance matrix for the non-zero power spectrum elements (i.e. those involving $\delta$ and $\gamma_E$) with high accuracy, a more sophisticated approach would be needed in order to obtain a precise estimate of the uncertainties for components involving $B$-modes. This is not a major concern, since $B$-mode power spectra are predominantly used as null tests, while cosmological parameter constraints are driven by the analysis of $\delta$ and $\gamma_E$. For completeness, Figure~\ref{fig:corr_diff_2bins} shows the difference between the correlation matrices for the sample covariance and the NKA estimator for all non-zero cross-correlations between different bins in the case of two redshift bins with different small-scale masks, where we find a similarly good agreement.

    \subsection{Quantitative comparison}\label{ssec:results.quant}
      \begin{figure}
        \centering
        \includegraphics[width=0.6\textwidth]{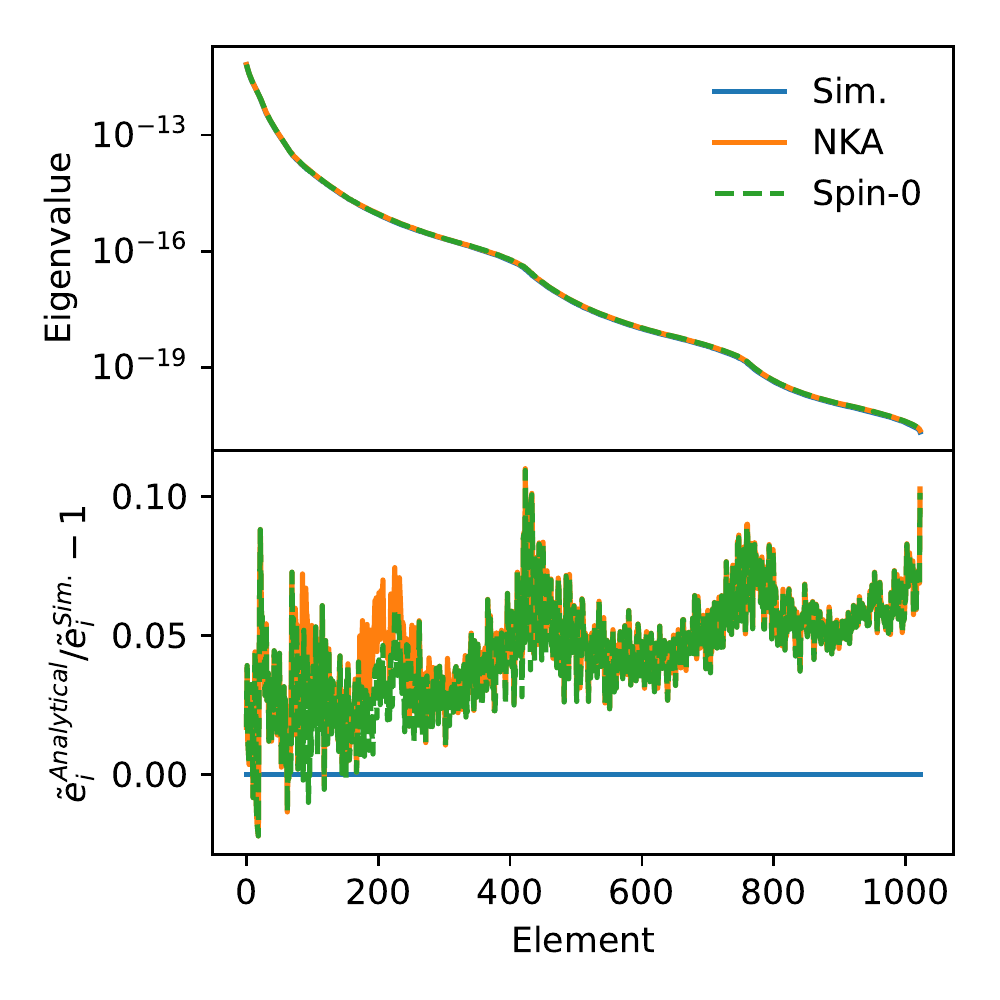}
        \caption{Eigenvalues of the single-bin covariance matrices for all power spectra involving $\delta$ and $\gamma_E$. Results are shown for the sample covariance (solid blue), the NKA estimator (solid orange) and the spin-0 approximation (dashed green). The NKA and spin-0 estimators are able to recover the covariance eigenvalues with an accuracy $\lesssim5\%$.} \label{fig:eigv_1bin}
      \end{figure}
      In order to quantify the validity of our analytical approximations, we need to compare the NKA covariance with the sample covariance estimated from the simulations. However, comparing two matrices is not as straightforward as comparing their elements one by one. The covariance between far-away bandpowers is expected to be very close to zero, and therefore a direct comparison of those elements would easily yield large relative differences simply due to the statistical noise in the sample covariance matrix. We will therefore quantify the differences between the different covariances making use of scalar quantities formed from them. The impact of the analytical approximations on the final parameter constraints will then be described in detail in Section \ref{ssec:results.cosmo}.
      
      As a first test to quantify the differences between covariance matrices, we compute the relative difference between their eigenvalues. This is shown in Figure~\ref{fig:eigv_1bin} for a data vector combining all auto- and cross-correlations between $\delta$ and $\gamma_E$ for a single redshift bin. The eigenvalues of both matrices are roughly similar, with relative differences of about $5\%$. The figure also shows the eigenvalues of the covariance matrix estimated using the Spin-0 approximation, which achieves a similar level of precision (even marginally higher in some cases).

      Another scalar quantity that can be used to compare different covariances is the $\chi^2$. For a given random data vector ${\bf d}$ with mean ${\bf m}$ and covariance matrix ${\rm Cov}$, this is given by
      \begin{equation}
        \chi^2=\left({\bf d}-{\bf m}\right)^T\cdot{\rm Cov}^{-1}\cdot\left({\bf d}-{\bf m}\right).
      \end{equation}
      \begin{figure}
        \centering
        \includegraphics[width=\textwidth]{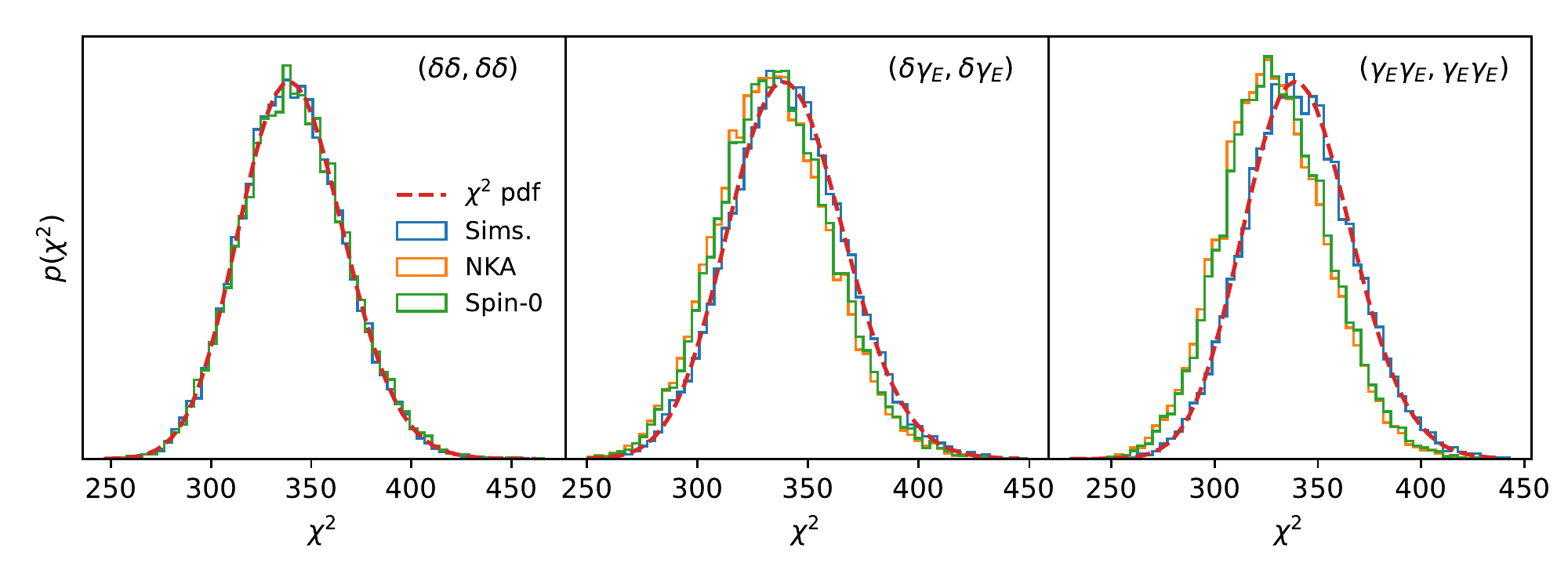}
        \caption{$\chi^2$ distributions for the cases without $\gamma_B$-modes. We compare the distributions obtained with the covariance matrix computed directly from the simulated power spectra (blue) and the two different analytical methods: NKA (orange) and spin-0 (green). In addition, the theoretical $\chi^2$ distribution has been included (dashed red). The distribution extracted from the simulation follows the theoretical expectation almost perfectly in all cases. In the cases of $\delta\times\gamma_E$ and $\gamma_E\times\gamma_E$, we observe small shifts in the peak $\chi^2$, of $\sim2$-$4\%$, while their width is recovered accurately (a difference $\lesssim 2\%$). We will show that these differences are due to the inaccuracy of the analytical approximations on the largest scales, and that they have a completely negligible impact on the final cosmological parameter constraints.} 
        \label{fig:chi2_1bin}
      \end{figure}
      \begin{figure}
        \centering
        \includegraphics[width=0.6\textwidth]{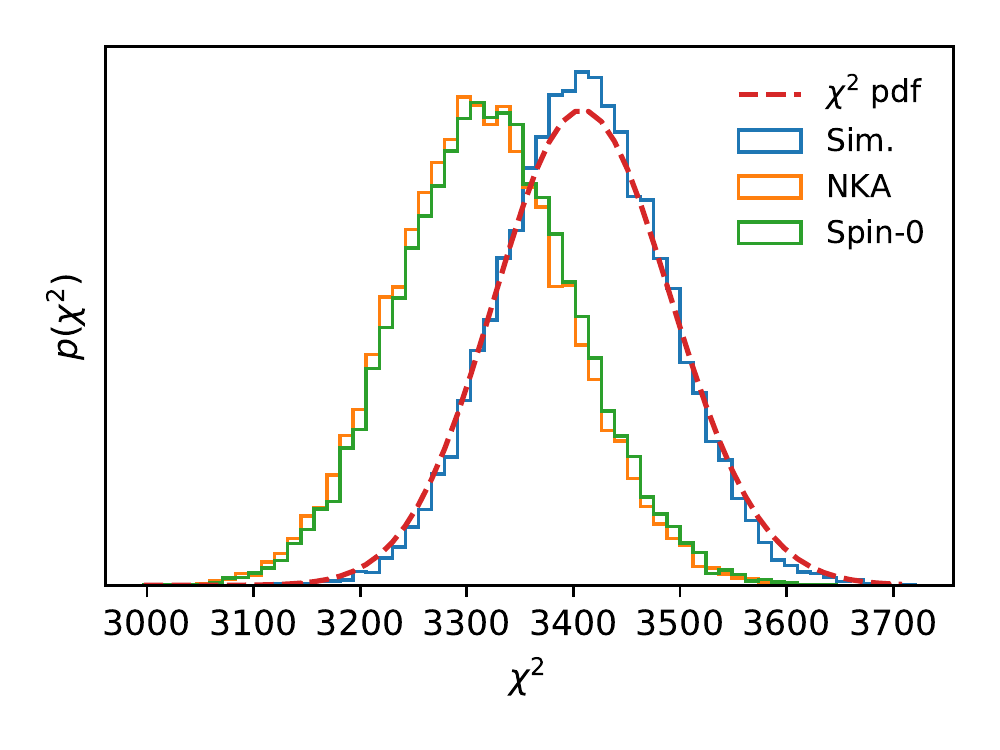}
        \caption{Same as Figure \ref{fig:chi2_1bin} for the combination of all correlations between $\delta$ and $\gamma_E$ for two redshift bins, where similar conclusions hold. In this case, the differences in the distribution means and widths are less than $3\%$ and $7\%$, respectively.}
        \label{fig:chi2_2bins}
      \end{figure}
      We compute this quantity for a data vector ${\bf d}$ composed of different auto- and cross-correlations for each of the 20,000 Gaussian simulations, with ${\bf m}$ given by the mean over all simulations and different choices of covariance matrix. Figure \ref{fig:chi2_1bin} shows the distribution of $\chi^2$ values for the three non-zero power spectra in the case of a single bin: $\delta$-$\delta$, $\delta$-$\gamma_E$ and $\gamma_E$-$\gamma_E$. The histograms show the distribution for the sample covariance matrix (blue), and the analytical NKA and Spin-0 estimators (orange and green, respectively). We additionally plot the theoretical $\chi^2$ distribution under the assumption that the underlying data vector is Gaussianly distributed (red dashed lines). In the simplest case of purely spin-0 quantities (leftmost panel), we find an excellent agreement between the different distributions. In the cases involving the spin-2 fields, we see noticeable differences between the distributions found with the sample covariance and the approximate ones. These differences are small, corresponding to less of a $2$ and $4\%$ shift in the mean $\chi^2$ for the $\delta$-$\gamma_E$ and $\gamma_E$-$\gamma_E$ cases, respectively, and a negligible variation in the width of the distributions. We therefore expect these differents to have a negligible effect on the posterior parameter distributions, as we show explicitly in Section \ref{ssec:results.cosmo}. The fact that these differences appear only for power spectra involving spin-2 quantities indicate that the NKA and spin-0 methods are imperfect at describing the additional mode coupling caused in the presence of a mask for higher-spin fields. This is rather obvious in the case of the spin-0 approximation but, interestingly, we find that the NKA and Spin-0 predictions yield results that are almost indistinguishable from each other. We therefore conclude that the additional approximation made in the NKA method for spin-2 fields -- neglecting the spatial derivatives of the mask -- is effectively equivalent to ignoring the spin nature of the fields involved. Note, however, that this is not the case for $B$-modes, where the NKA estimator outperforms the spin-0 approximation by up to one order of magnitude, even though its accuracy is very poor (as we described in the previous section). For completeness, Figure \ref{fig:chi2_2bins} shows the distribution of $\chi^2$ values for a data vector composed of all possible auto- and cross-correlations of $\delta$ and $\gamma_E$ for the case of two redshift bins, where similar conclusions hold.

\begin{figure}
        \centering
        \includegraphics[width=\textwidth]{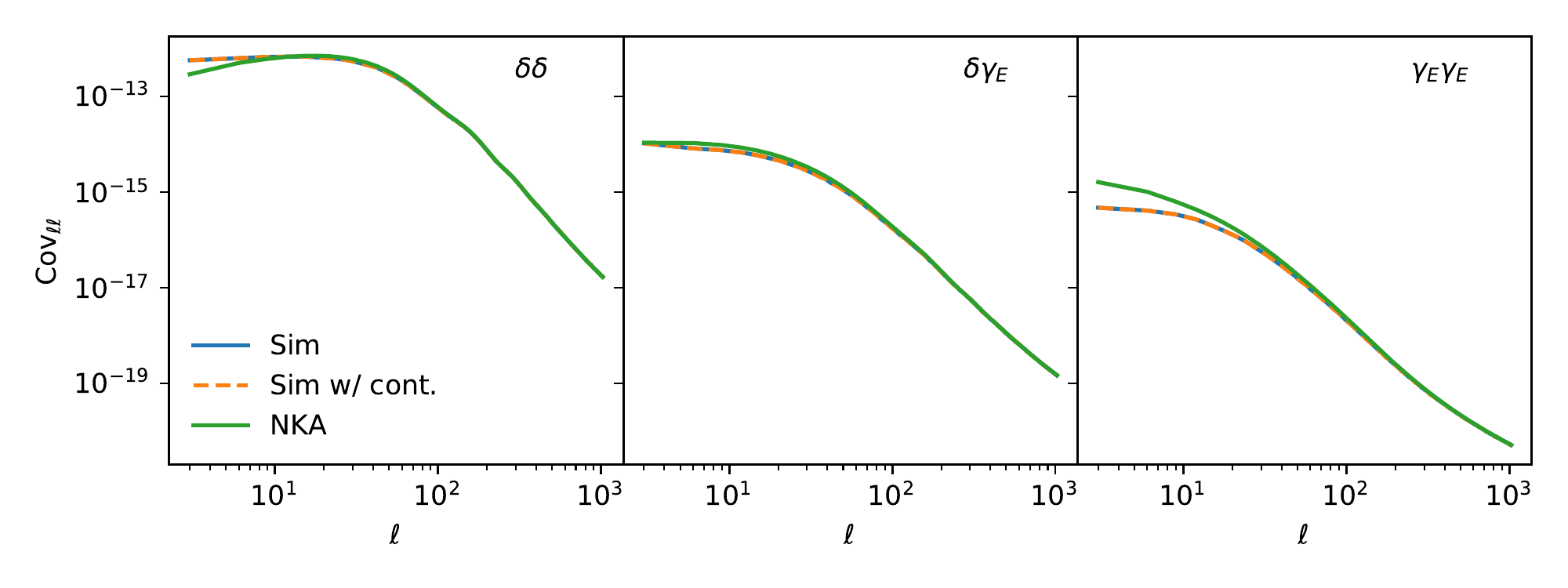}
        \caption{Diagonal of the covariance matrix estimated from simulations without contaminants (solid blue), with 200 contaminants deprojected (dashed orange) and the NKA estimator (solid green). The differences between the simulated cases are significantly smaller than those associated with the NKA estimator. The NKA method is therefore equally valid to approximate the covariance even in the presence of contaminant deprojection.}
        \label{fig:diag_conts}
      \end{figure}

      We have also explored the impact of contaminant deprojection on the different covariance matrix estimates. The loss of modes due to deprojection can potentially increase the variance of the power spectrum estimates, affecting the accuracy with which an analytical estimator would be able to recover the covariance. Figure \ref{fig:diag_conts} shows the diagonal of the covariance matrix for the $\delta$-$\delta$,      $\delta$-$\gamma_E$ and $\gamma_E$-$\gamma_E$ power spectra. Each panel displays the diagonal for the sample covariance matrix estimated from simulations without contaminants or contaminant deprojection (blue
      line), the sample covariance from simulations with contaminants and contaminant deprojection (dashed orange line) and for the NKA covariance   (green line). We see that the power spectrum uncertainties are almost indistinguishable with or without deprojection, and that those relative differences are much smaller than the differences between the sample covariance and the NKA estimator. We therefore conclude that, except in the case where a very large set of contaminant maps are deprojected (comparable with the number of unmasked pixels in the map), the analytical approximation to the covariance matrix should be as accurate as in the absence of contaminants (i.e. accurate enough).

      Figure \ref{fig:diag_conts} serves also to illustrate another important
      point. In all our tests we find that the largest differences between the
      sample and NKA covariances occur on large scales. For $\delta$-$\delta$
      correlations, the effect is limited to the first few multipoles
      ($\ell\lesssim10$), while for $\gamma_E$-$\gamma_E$ we are only able to
      recover the sample covariance errors within $5\%$ for $\ell\gtrsim40$.
      This is the main source of the small mismatch observed in
      Figure~\ref{fig:chi2_1bin}. Note, however, that most of the cosmological
      information is obtained from the higher multipoles, due to their higher
      statistical weight, and, as we will show in the next section, the effect
      on the final parameter constraints is negligible. If accurate
      covariance matrix elements are needed on these large scales for cosmic shear, they
      can be estimated alternatively making use of fast, low resolution
      simulations (e.g. {\tt HEALPix} $N_{\rm side}=64$ maps), or computed
      exactly as in Ref.~\cite{Efstathiou:2006eb,2017A&A...602A..41C}.

    \subsection{Impact on parameter estimation}\label{ssec:results.cosmo}
      \begin{figure}
        \centering
        \includegraphics[width=0.6\textwidth]{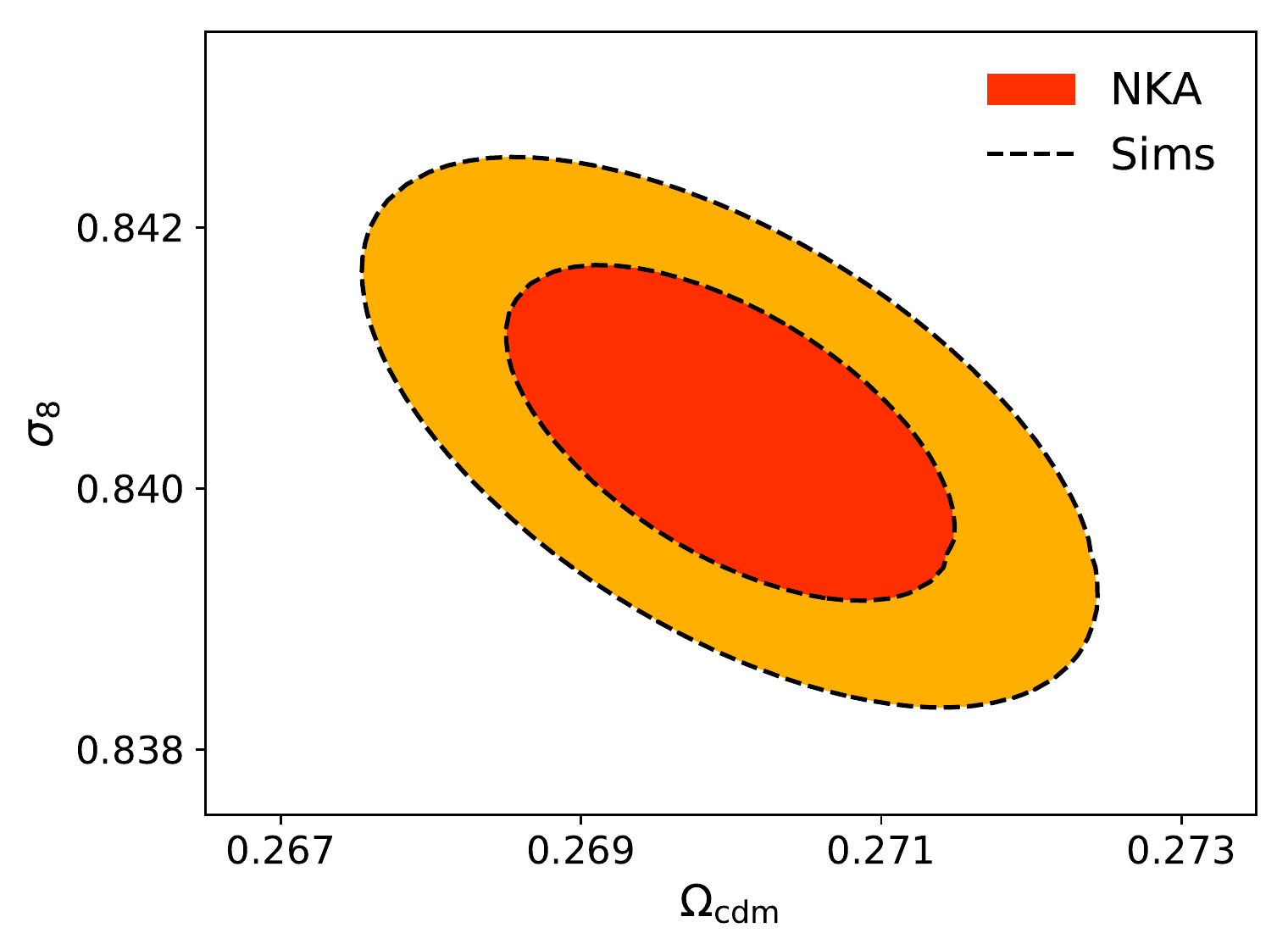}
        \caption{1$\sigma$ and 2$\sigma$ contours for $\Omega_M$ and $\sigma_8$ found from two different estimates of the power spectrum likelihood for a dataset containing all possible cross-power spectra between $\delta$ and $\gamma_E$ for two redshift bins. The filled contours correspond to the parameter likelihood evaluated using the NKA estimator for the power spectrum covariance matrix, while the dashed black contours correspond to the same calculation done with the sample covariance matrix corrected by the factor in Eq. \ref{eq:hartlap}. In both cases we assume a Gaussian likelihood. We find a remarkably good agreement between both likelihoods, highlighting the negligible impact of the approximations involved in the NKA method on the final parameter estimates.} \label{fig:contours}
      \end{figure}
      Ultimately, the most important test to judge the accuracy of the
      analytical covariance matrix estimators implemented here is to study
      their impact on the posterior distribution of cosmological parameters
      derived from power spectrum measurements. Assuming flat priors and
      a fiducial Gaussian likelihood approximation~\cite{2008PhRvD..77j3013H}, in which the covariance matrix is computed only once for the fiducial model, the posterior distribution for parameters $\vec{\theta}$ is simply given by:
      \begin{equation}\label{eq:parpos}
        -2\log p(\vec{\theta}|\hat{\bf C})=\left(\hat{\bf C}-{\bf C}(\vec{\theta})\right)^T\cdot{\rm Cov}^{-1}\cdot\left(\hat{\bf C}-{\bf C}(\vec{\theta})\right)+{\rm const.},
      \end{equation}
      where $\hat{\bf C}$ is a vector of power spectrum measurements, ${\rm Cov}$ is their covariance matrix, and ${\bf C}(\vec{\theta})$ is their theoretical prediction for parameters $\vec{\theta}$. It is worth noting that this Gaussian likelihood is not accurate on large scales, where the small number of modes invalidates the application of the central limit theorem. Even in this regime, the likelihood can be approximated through the method described in \cite{2008PhRvD..77j3013H}, which still requires an accurate estimate of the power spectrum covariance.

      We explore $\log p(\vec{\theta}|\hat{\bf C})$ for the two parameters $\vec{\theta}\equiv(\Omega_M,\sigma_8)$, for a data vector composed of all possible auto- and cross-correlations between $\delta$ and $\gamma_E$ in the case of two redshift bins described in Section \ref{ssec:results.sims}. In this simple two-dimensional scenario, we simply sample the distribution in a regular grid of 100 by 100 points for each parameter. We construct a data vector $\hat{\bf C}$ from the theoretical prediction for the experimental setup described in Section \ref{ssec:results.sims}, and produce theoretical predictions for it at each grid point using {\tt CCL}. Note that, when evaluating the posterior for the sample covariance matrix, one needs to correct for the finite number of simulations used to construct the covariance. In most situations this can be done simply by rescaling the inverse covariance matrix by a factor given by \cite{2007A&A...464..399H}
      \begin{equation}\label{eq:hartlap}
        {\rm Cov}^{-1}\,\longrightarrow \frac{N_s-2-N_{\rm data}}{N_s-1}\,{\rm Cov}^{-1},
      \end{equation}
      where $N_s=20,000$ is the number of samples used to estimate the
      covariance, and $N_{\rm data}=3510$ is the number of data points.
      
      The 68\% and 95\% confidence level contours associated with the posterior distributions for the sample covariance and the NKA estimator are shown in Figure~\ref{fig:contours}. We find that both distributions agree with each other remarkably well, and that the $1\sigma$ errors for each parameter agree for both covariances up to $0.3\%$. Note that, since we have not included any statistical noise in the data vector $\hat{\bf C}$, the relative difference in the means of both distributions is zero by construction (since Eq. \ref{eq:parpos} is bounded from below by zero). When adding Gaussian statistical noise compatible with the sample covariance matrix, we observe small differences (smaller than 0.3$\sigma$) in the best fit parameters found with both covariance matrices. These differences, however, are not systematic, and we have verified that averaging over several noise realizations does not yield biased best-fit parameters.
      
      We therefore conclude that the analytical approximations for the power spectrum covariance matrix explored here are able to reproduce the true posterior distribution of cosmological parameters to very high accuracy.

  \section{Discussion}\label{sec:discussion}
    Estimating accurate covariance matrices for projected two-point
    correlators is an ubiquitous problem in modern cosmology
    \cite{2013PhRvD..88f3537D,2013MNRAS.432.1928T,2017arXiv170609359K}, but it
    is particularly relevant for large-scale structure datasets. This problem
    is further complicated in this case, in comparison with e.g. CMB
    experiments, by two factors: 
    the fact that the fields involved are
    non-Gaussian at some level and the arguably higher complexity of the sky masks used in optical datasets.
    The main impact of survey geometry is the statistical
    coupling it induces between different Fourier/harmonic modes, which must
    be accurately characterized in order to obtain reliable estimates of the
    posterior distribution for cosmological parameters. In this paper we have
    focused on the impact of survey geometry on the dominant Gaussian (i.e.
    disconnected) part of the covariance matrix.
    
    We have described and generalized existing analytical approaches to
    estimate the covariance matrix for pseudo-$C_\ell$ power spectrum
    estimators
    \cite{2004MNRAS.349..603E,2005MNRAS.360.1262B,2017A&A...602A..41C}, and
    implemented them in the public code {\tt NaMaster}
    \cite{2019MNRAS.484.4127A}, making it straightforward to fully account for
    the effects of survey geometry on the data uncertainties. With these
    approximations, computationally speaking, the problem of estimating a
    covariance matrix is as complex as that of computing the power spectrum
    itself, and scales with the number of pixels in the map as $N_{\rm
      pix}^{3/2}$. We leave for future work the study of the impact of
      extra complications, as position-dependent noise, on their performance.
    
    The main finding of this paper is the excellent performance of the
    analytical methods described in Section \ref{ssec:theory.pclcov}. We have
    shown that
    the NKA estimator (see Section \ref{ssec:theory.approx}) is able to
    recover the covariance matrix for all power spectra with non-zero signal
    expectation value (i.e. those involving the galaxy overdensity $\delta$ or
    the $E$-mode shear $\gamma_E$), as well as the posterior distribution for
    cosmological parameters, to a high degree of accuracy.

    More in detail, we have also found that the impact of contaminant
    deprojection on the covariance matrix, through the corresponding loss of
    modes, is negligible unless a very large number of contaminant templates
    are removed. This simplifies the procedure to estimate covariance matrices
    for galaxy clustering data, which are particularly sensitive to a large
    number of astrophysical and observational systematics. Additionally, we
    have found that, although the NKA estimator is accurate enough, it is not
    able to perfectly capture the additional effects of mode coupling that are
    present for spin-2 fields, and that a simpler approach treating the shear
    $E$ modes as a spin-0 object is able to reach similar levels of accuracy.
    Due to the imperfect treatment of the $E/B$ mixing caused by the sky mask
    in the NKA estimator, we also find that the predicted covariance matrix
    for any power spectra involving $B$ modes differs significantly from the
    true sample covariance, and, therefore, this approach cannot be used to
    reliably estimate the uncertainties of $B$-mode power spectra. Likewise,
    we find that the NKA estimator yields inaccurate estimates of the power
    spectrum uncertainties on the largest scales ($\ell\lesssim50$). Although
    these modes carry a substantially smaller statistical weight, if more
    accurate covariances are needed on these large scales, they can be easily
    computed making use of fast low-resolution simulations or exactly
    following Refs.~\cite{Efstathiou:2006eb,2017A&A...602A..41C}.
    
    In spite of these shortcomings, we find that the approximations described in this paper are able to provide estimates of the power spectrum covariance matrix that are sufficiently accurate for current and future tomographic large-scale structure cosmological datasets. The main advantage of this approach is the computational cost, which is comparable to that of estimating the power spectrum in the first place, and which  scales with pixel resolution in a similar way. This is, therefore, significantly less time-consuming than generating large numbers of mock datasets (even simple Gaussian or log-normal realizations), and more reliable than the traditional jackknife resampling techniques. The method, in all its generality, is currently implemented in the public code {\tt NaMaster}. Future extensions to this work will focus on improving the estimator for power spectra involving $B$ modes, and incorporating the impact of $E/B$-mode purification \cite{2002PhRvD..65b3505L,2003PhRvD..67b3501B,2006PhRvD..74h3002S}.
    
  \acknowledgments
  We thank Thibaut Louis, Eva-Maria Mueller, Francisco Javier S\'anchez and
  An\v ze Slosar for useful comments and discussion. CGG is supported the
  Spanish grant BES-2016-077038, partially funded by the ESF and by
  AYA2015-67854-P from the Ministry of Industry, Science and Innovation of
  Spain and the FEDER funds. He was partially supported by a Balzan Fellowship
  while in Oxford. He would like to thank New College and the Department of
  Physics at Oxford for their hospitality. DA acknowledges support from STFC
  through an Ernest Rutherford Fellowship, grant reference ST/P004474/1. EB
  acknowledges support from the Beecroft Trust.  Some of the results in this
  paper have been derived using the HEALPix~\cite{Gorski:2004by} package.
  
  \appendix
  \section{Flat sky}\label{app:flat}
    \begin{figure}
      \centering
      \includegraphics[width=0.6\textwidth]{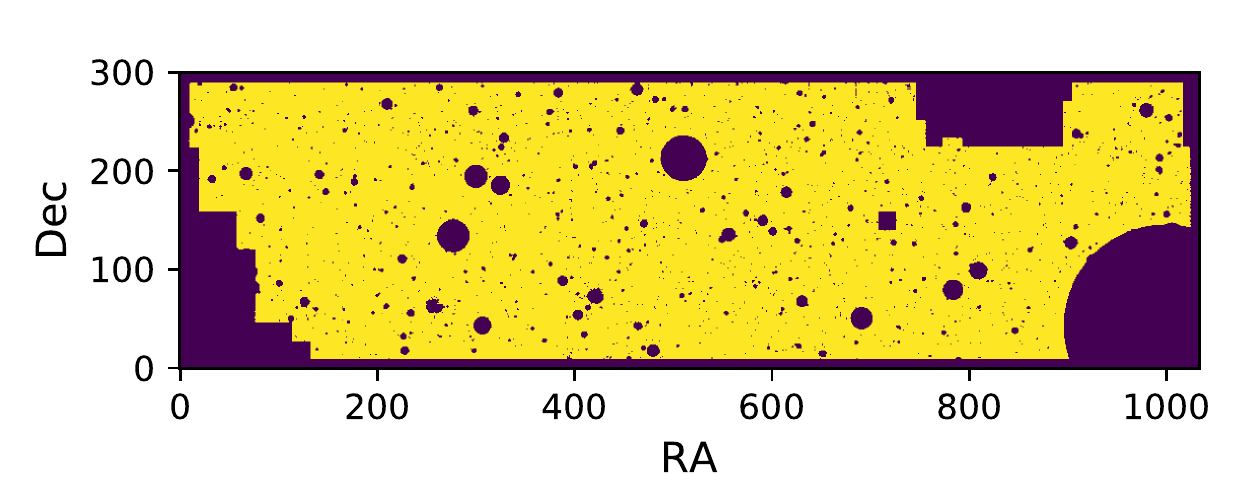}
      \caption{Flat-sky mask used in the study of the analytical methods presented in this work in the flat-sky regime. The mask was constructed from the bright-object mask distributed with the first data release of the HSC collaboration \cite{2018PASJ...70S...4A} for the VVDS field. As in the curved-sky case, it is used both for galaxy clustering and shear for simplicity.} \label{fig:mask_flat}
    \end{figure}

    \begin{figure}
      \centering
      \includegraphics[width=\textwidth]{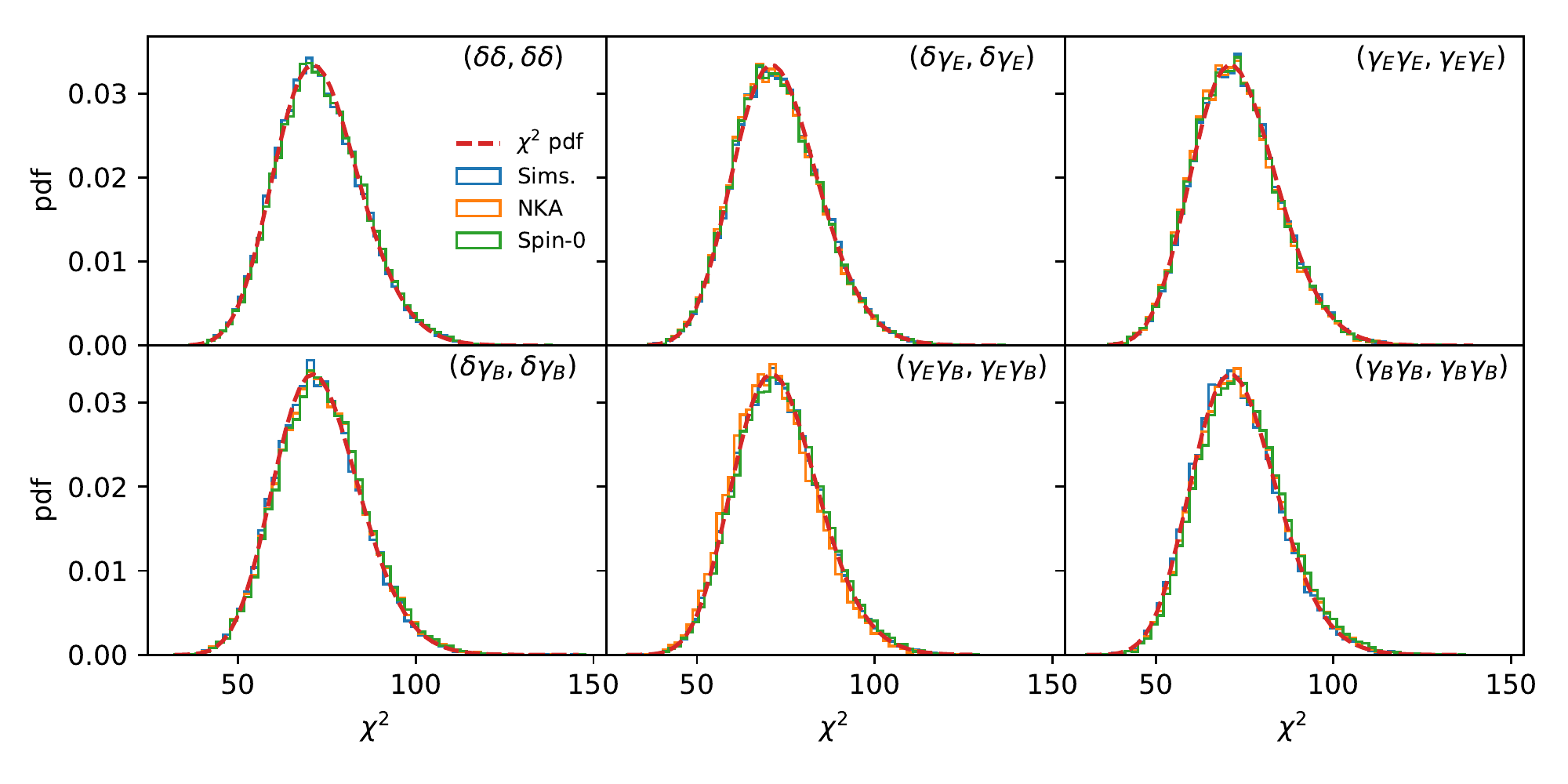}
      \caption{Same as Figure \ref{fig:chi2_1bin} for flat-sky fields. This figure additionally shows the $\chi^2$ distributions for power spectra involving $B$-modes. For the smaller scales covered by these flat-sky simulations, the NKA and spin-0 estimators work remarkably well, even for $B$-mode spectra. The distribution mean shift and width difference is of about $\lesssim 2\%$ for all cases.}
      \label{fig:chi2_1bin_flat}
    \end{figure}
    We have repeated the analysis described in Section \ref{sec:results} on flat-sky realizations, making use of the flat-sky implementation of {\tt NaMaster}. We generate Gaussian realizations of the galaxy overdensity and shear maps making use of flat-sky extensions of the methods described in Section \ref{ssec:results.sims}. In this case we use a high-resolution mask constructed from the bright-object mask distributed with the first data release of the HSC collaboration \cite{2018PASJ...70S...4A} for the VVDS field, which is shown in Figure~\ref{fig:mask_flat}. 
    
    We find similar levels of accuracy in the NKA and spin-0 estimators compared to the curved-sky case. The higher resolution and smaller area of these simulations allow us to focus on the small-scale galaxy clustering and lensing power spectra, covering the range of multipoles $\ell\in(120,17640)$ in constant bandpowers of width $\Delta \ell=240$. On these small scales, the shear power spectrum is more dominated by noise (e.g. see Fig. \ref{fig:cl-2bins}), and therefore there is roughly the same power in $E$ and $B$ modes. This reduces the sensitivity of the method to an inaccurate treatment of $E/B$ leakage, and the agreement between the sample covariance matrix and the NKA estimator improves significantly. This can be seen in Figure~\ref{fig:chi2_1bin_flat}, which shows the $\chi^2$ distributions for all possible power spectra (including those including $B$ modes) in the case of a single redshift bin. In all cases we find a good agreement between the $\chi^2$ distributions derived from all covariance matrix estimates (sample covariance, NKA and spin-0 in blue, orange and green respectively), which also accurately follow the expected $\chi^2$ distribution for the corresponding number of degrees of freedom (red dashed line).
    
  \section{Software implementation}\label{app:namaster}
    In addition to the code functionality described in Section 3 of \cite{2019MNRAS.484.4127A}, we have now included the capability to estimate Gaussian covariance matrices using the NKA method. This functionality is structured around a {\tt python} class called {\tt NmtCovarianceWorkspace}. These objects are used to compute and store the covariance mode-coupling coefficients in Eqs. \ref{eq:coeff_cmcm} and \ref{eq:coeff_cmcm_flat}. They are initialized from two pairs of fields corresponding to the two power spectra for which the covariance is required. Once initialized, these coefficients can be reused for any other set of fields with the same combination of sky masks. {\tt NaMaster} then provides routines to estimate covariance matrix elements making use of the coupling coefficients stored in a {\tt NmtCovarianceWorkspace} object and best-guess power spectra for the fields involved using Eqs. \ref{eq:cov_sph} and \ref{eq:cov_flat}. Further details about the implementation and practical examples can be found in \url{https://namaster.readthedocs.io/en/latest/sample_covariance.html}.

  \bibliographystyle{JHEP}
  \bibliography{paper}

\end{document}